# Underwater Source Localization Using TDOA and FDOA Measurements With Unknown Propagation Speed and Sensor Parameter Errors

**BINGBING ZHANG[1], (Student Member, IEEE), YONGCHANG HU[2], (Member, IEEE), HONGYI WANG[1], AND ZHAOWEN ZHUANG[1]**

[1]School of Electronic Science, National University of Defense Technology, Changsha 410073, China
[2]Faculty of Electrical Engineering, Mathematics and Computer Science, Delft University of Technology, 2624AL Delft, The Netherlands

Corresponding author: Bingbing Zhang (zbbzb@nudt.edu.cn)

This work was supported in part by the Hunan Science and Technology Project under Grant 2016JC2008, in part by the China Postdoctoral Science Foundation under Grant 2016T90978, and in part by the National Nature Science Foundation of China under Grant 61604176.

**ABSTRACT** Underwater source localization problems are complicated and challenging: 1) the sound propagation speed is often unknown and the unpredictable ocean current might lead to the uncertainties of sensor parameters (i.e., position and velocity); 2) the underwater acoustic signal travels much slower than the radio one in terrestrial environments, thus resulting into a significantly severe Doppler effect; and 3) energy-efficient techniques are urgently required and hence in favor of the design with a low computational complexity. Considering these issues, we propose a simple and efficient underwater source localization approach based on the time difference of arrival and frequency difference of arrival measurements, which copes with unknown propagation speed and sensor parameter errors. The proposed method mitigates the impact of the Doppler effect for accurately inferring the source parameters (i.e., position and velocity). The Cramér–Rao lower bounds (CRLBs) for this kind of localization are derived and, moreover, the analytical study shows that our method can yield the performance that is very close to the CRLB, particularly under small noise. The numerical results not only confirm the above conclusions but also show that our method outperforms other competing approaches.

**INDEX TERMS** Underwater localization, algebraic solution, sound propagation speed uncertainty, sensor node uncertainty, time difference of arrival (TDOA), frequency difference of arrival (FDOA).

## I. INTRODUCTION

Location-awareness is an important and indispensable feature for a variety of underwater scenarios, such as ocean resource exploration, environment monitoring, disaster prevention and underwater navigation [1]. Sensors and/or vehicles, which include autonomous underwater vehicles (AUVs), underwater gliders, remotely operated underwater vehicles (ROVs) and underwater buoys and etc., particularly value and hence often require their location information to be associated with the collected data or for navigation purpose. Moreover, for enhancing the communication and networking performance, the location information is also very useful for improving techniques like topology control, routing and packet collision avoidance [2]. Therefore, underwater source localization becomes a rather prevalent and popular topic in the recent years. In this paper, any underwater device that transmits acoustic signals can be viewed as a source, and the ultimate goal of our work is accurately inferring the source parameters (i.e. position and velocity) in realistic underwater environments.

Obviously, the Global Positioning System (GPS) is infeasible for underwater scenarios due to the different transmission medium. Though it can still provide some reference receivers that are set afloat, e.g. buoys, with useful information like positions and velocities. As for those underwater, we also assume their positions and velocities are known *a priori* to the localization phase. In a nutshell, all the reference nodes, which will be referred to as sensor nodes (SNs) later, are of necessary assistance to locating the underwater source. For signalling, we adopt the acoustic signal, which is widely used for underwater scenarios. This is because the sound wave can propagate several kilometres and cover a large ocean area, whereas the radio-frequency signal decays very fast in the water and can only spread within a short distance. Commonly used measurements for source localization include time-of-arrive (TOA) [3], time-difference-of-arrival (TDOA) [4], angle-of-arrive (AOA) [5], received signal strength (RSS) [6] and differential RSS (DRSS) [7]. Nevertheless, employing the AOA requires antenna array, which is often too expensive or difficult in practice. Detecting the energy of the received signal is rather hard due to the unpredictable variations in underwater channel [8], hence the RSS or DRSS measurement is still barely considered for underwater source localization. The time measurements (the TOA and the TDOA) are very favourable for underwater source localization. Seeing that the use of the TOA heavily relies on the clock synchronization between both the source-SN and the SN-SN links while the requirements for the TDOA is much less strict (only among the SNs), we mainly consider the latter in our work for practical convenience. Another reason that we prefer the TDOA is that, in some uncooperative scenarios, the signals radiated from the source mostly do not carry the time stamp, thus making the clock synchronization between the source and the SNs even more difficult. In addition to the TDOA, we also exploit another kind of measurement for localization. In realistic underwater scenarios, even when nodes are not travelling, they can barely keep static due to the unpredictable ocean currents. Therefore, we have to consider the Doppler effect. Fortunately, the Doppler shifts can be collected at signal receivers, from which we can further obtain the frequency-difference-of-arrival (FDOA) measurements [9]. Fusing the TDOA and the FDOA not only is helpful for improving localization performance but also provides the useful velocity estimates [10]. In a nutshell, we base our proposed method on the TDOA/FDOA measurement set in this paper.

The use of the measurement fusion will undoubtedly lead to a severer non-linearity issue, resulting into a more strenuous localization problem. In our work, what makes it even worse is the harsh underwater environment. First, the SN parameters (i.e. position and velocity) information might be inaccurate. In real-life, the SNs are mounted on the seabed, suspended in the water column or afloat on the sea surface [11], as depicted in Fig. 1. All of them, although the afloat ones can acquire their location information from the GPS system, are always drifting around with the unpredictable ocean currents that vary significantly over time and space [12], thus making the obtained location information not very accurate. Particularly, for those in deep water, apart from the immense implementation cost, their location knowledge might not even be guaranteed *a priori*. Apparently, if we ignore the imprecision of SN parameters information, the localization performance will severely deteriorate [13]. This strongly motivates us to take the SN parameter errors into account in this paper. Second, the fact that the acoustic signalling is considered for underwater scenarios implies a totally unpredictable sound propagation speed. It has been shown that the sound speed profile (SSP) is subject to temperature, pressure, salinity and depth in underwater environment [14], [15], which means the SSP is basically impossibly known *a priori*. This has also been observed in other underwater localization literature [16]–[19]. Note that the travel time between a transmitter and a receiver can always be converted into a slant range by multiplying an effective sound speed. In our work, we cope with this issue by viewing the unknown sound speed as an unknown nuisance parameter that needs to be jointly estimated.

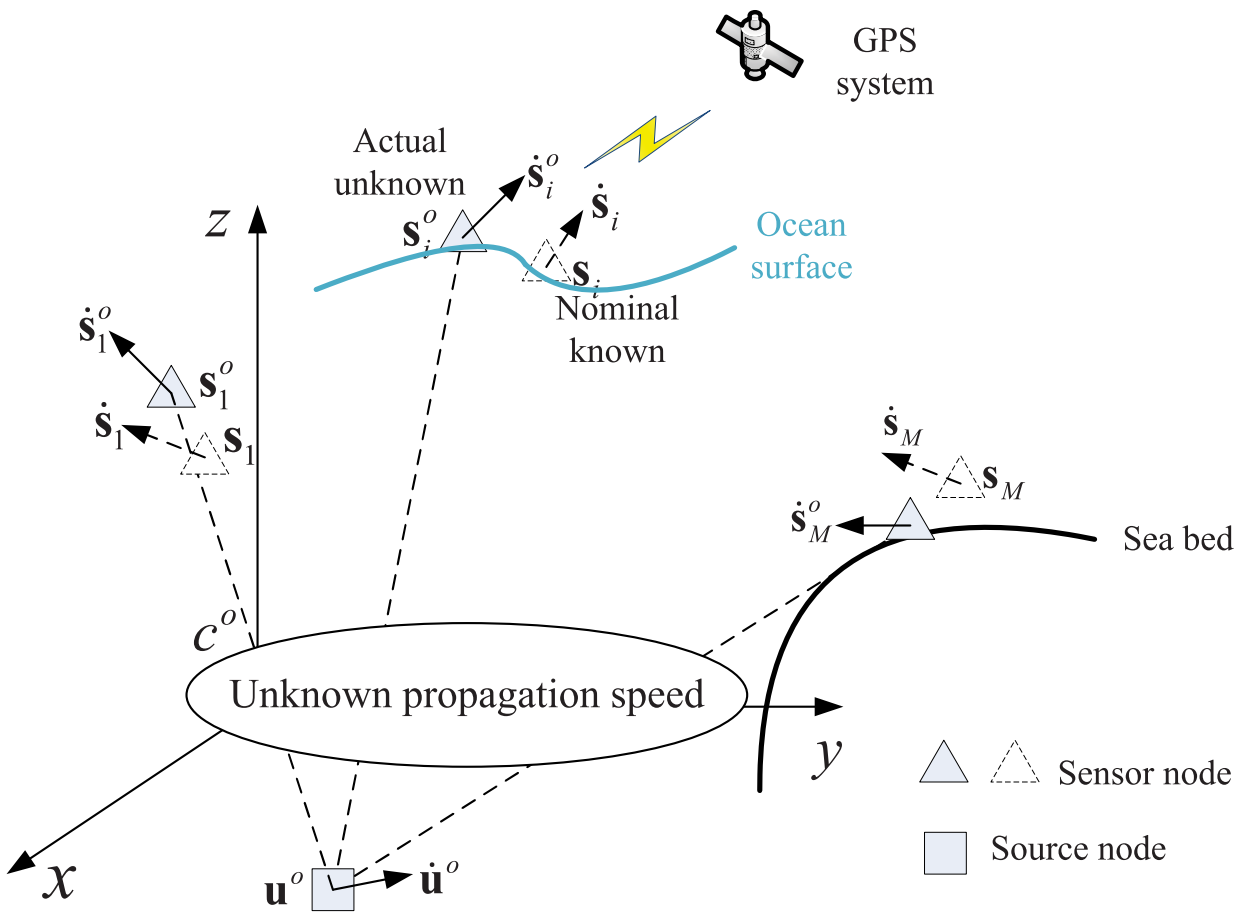

**FIGURE 1.** Underwater localization scenario.

There are techniques from estimation theory that can be applied to the localization problem, such as least squares (LS) estimator, convex optimization and compressed sensing (CS) [20], [21]. Specifically, existing methods for source localization using the TDOA/FDOA measurements can be divided into three categories: Taylor-series expansion based [22], [23], semi-definite programming (SDP) based [24]–[26] and LS based [10], [13], [27], [28]. The first kind of methods use the Taylor-series expansion to linearize the optimization problem in the vicinity of position and velocity estimates that should be close to the actual values. Note that selecting appropriate initialization values is practically difficult but very important, otherwise we might obtain a local solution. The SDP based techniques relax the original non-convex problem onto a convex set such that the new optimization problem can be efficiently solved [29]. However, for a good estimation accuracy, the relaxation needs to be very tight, which is rather challenging, not to mention the high computational complexity for solving the final problem. Without any of those disadvantages, the LS based methods are very favourable for solving localization problems. In [10], the famous two-step weighted LS (TS-WLS) method was proposed, where the first step provides initial position and velocity estimates, which will be fine-tuned later in the second step. A robust version of TS-WLS method was also reported in [13], where the SN position and velocity uncertainties were taken into account. M. sun, et al. further extended this kind of method in case of multiple disjoint sources in [27]. A more recent improved TS-WLS solution was introduced in [28], where the second step is designed with a better fine-tuning procedure.

Nonetheless, all the aforementioned methods assume a known sound propagation speed, which is very impractical and unrealistic. Admittedly, there indeed exist some other literature that consider an unknown sound speed. In [30], the TS-WLS method was developed into a three-step algorithm for jointly estimating the location and the unknown speed, however the Doppler effect is ignored and only the TDOA measurements are used therein. More importantly, the authors assume the actual SN locations to be exactly known *a priori*, which might not be suitable for underwater scenarios. In [31], an SDP localization approach using the TOA measurements was devised in the presence of both sensor position and speed uncertainties. As already mentioned, this method will certainly suffer from the clock synchronization constraint and the high computational complexity. An efficient algebraic localization technique was also proposed coping with SN position and speed uncertainties [32], though it is designed for multistatic sonar scenarios with a totally different measurement model. In a nutshell, the study for realistic underwater source localization is still in its infancy. To the best of our knowledge, in current literature, there does not exist any algebraic localization approach using the TDOA/FDOA measurements and coping with SN parameters and sound propagation speed uncertainties.

Therefore, we would like to enrich the study of this field by proposing a new efficient solution for realistic harsh underwater environments. In this paper, we assume a totally unknown sound propagation speed and erroneous SN parameters and, based on the TDOA/FDOA measurement set, the introduced method can still jointly estimate the location and velocity of a moving source with a notably good performance. We also derive and study the corresponding Cramér-Rao lower bound (CRLB). Numerical simulations have also been conducted for evaluating the localization performance of our proposed method.

The rest of this paper is organized as follows. Section II presents the underwater localization problem in this paper. Then, Section III derives and studies the CRLB for the considered localization problem. The impact of the unknown sound propagation speed is particularly discussed. Next, our new efficient underwater localization approach is introduced in Section IV and we also analytically study the performance of the proposed method in Section V. The numerical results in Section VI not only support the previous conclusions but also show that our method outperforms other competing approaches. Finally, Section VII summarizes this paper.

### Notation

The following notations are used in this paper. $\mathbb{R}^3$ indicates an 3-dimensional real space; upper (lower) bold-face letters stand for matrices (vectors); superscript $T$ denotes the transpose of a matrix (vector); $(*)^o$ indicates the actual value of $(*)$; $sign(*)$ denotes the signum function; The operators $\odot$, $\otimes$ and $./$ designate the element-wise product, Kronecker product and element-wise division, respectively; $||\cdot||$ is the Euclidean distance norm and $\mathbf{a}(p:q)$ is a subvector formed by the $p$-th to the $q$-th element of vector $\mathbf{a}$; $[\mathbf{X}]_{m,n}$ represents the element on the $m$th row and $n$th column of the matrix $\mathbf{X}$; $\mathbf{1}$ and $\mathbf{0}$ are vectors of 1 and 0; $\mathbf{I}$ denotes the identity matrix (size indicated in the subscript if necessary); $\mathbb{P}(\mathbf{x}|\mathbf{x}^o, \mathbf{Q_x})$ indicates the Gaussian distribution with expectation $\mathbf{x}^o$ and covariance matrix $\mathbf{Q_x}$; diag$(*)$ is the diagonal matrix with the elements of $*$ on its diagonal and Tr$(*)$ denotes the trace of a square matrix $*$; blkdiag$(\mathbf{A}, \mathbf{B})$ is a matrix with $\mathbf{A}$ and $\mathbf{B}$ on its diagonal and all other elements zero.

## II. UNDERWATER LOCALIZATION PROBLEM

In this section, we formulate the underwater localization problem used throughout this paper. First, we assume a single moving source located at $\mathbf{u}^o = [x^o, y^o, z^o]^T \in \mathbb{R}^3$ with velocity $\dot{\mathbf{u}}^o = [\dot{x}^o, \dot{y}^o, \dot{z}^o]^T$ and $M$ SNs located at $\mathbf{s}_i^o = \left[x_i^o, y_i^o, z_i^o\right]^T \in \mathbb{R}^3$ with velocities $\dot{\mathbf{s}}_i^o = \left[\dot{x}_i^o, \dot{y}_i^o, \dot{z}_i^o\right]^T$, $i = 1, 2, \ldots, M$ in an underwater environment, as depicted in Fig. 1. Stacking the actual SN parameters, i.e., locations $\mathbf{s}_i^o$ and velocities $\dot{\mathbf{s}}_i^o$, into a single vector, we obtain the actual SN parameter vector $\boldsymbol{\beta}^o \triangleq \left[(\mathbf{s}^o)^T, (\dot{\mathbf{s}}^o)^T\right]^T$, where $\mathbf{s}^o \triangleq \left[(\mathbf{s}_1^o)^T, (\mathbf{s}_2^o)^T, \ldots, (\mathbf{s}_M^o)^T\right]^T$ and $\dot{\mathbf{s}}^o \triangleq \left[(\dot{\mathbf{s}}_1^o)^T, (\dot{\mathbf{s}}_2^o)^T, \ldots, (\dot{\mathbf{s}}_M^o)^T\right]^T$. However, in practice, the known SN parameters $\boldsymbol{\beta}$ are subject to errors and hence we would like to refer $\boldsymbol{\beta}$ to the nominal values of $\boldsymbol{\beta}^o$ as

$$\boldsymbol{\beta} = \boldsymbol{\beta}^o + \Delta\boldsymbol{\beta},$$

where the errors $\Delta\boldsymbol{\beta} \triangleq \left[\Delta\mathbf{s}^T, \Delta\dot{\mathbf{s}}^T\right]^T$ are assumed to be zero-mean Gaussian distributed with covariance matrix $\mathbf{Q}_{\boldsymbol{\beta}} = E\left[\Delta\boldsymbol{\beta}\Delta\boldsymbol{\beta}^T\right]$. Note that

$$\Delta\mathbf{s} = \mathbf{s} - \mathbf{s}^o = \left[\Delta\mathbf{s}_1{}^T, \Delta\mathbf{s}_2{}^T, \ldots, \Delta\mathbf{s}_M{}^T\right]^T$$

and

$$\Delta\dot{\mathbf{s}} = \dot{\mathbf{s}} - \dot{\mathbf{s}}^o = \left[\Delta\dot{\mathbf{s}}_1^T, \Delta\dot{\mathbf{s}}_2^T, \ldots, \Delta\dot{\mathbf{s}}_M^T\right]^T.$$

Then, denoting the distance between the source node and the $i$-th SN as

$$r_i^o = \left\|\mathbf{u}^o - \mathbf{s}_i^o\right\|, \tag{1}$$

we select the first SN to be the reference node and express the TDOA measurement set as

$$t_{i1} = \frac{1}{c^o} r_{i1}^o + n_{i1}, \quad i = 2, 3, \ldots, M, \tag{2}$$

where $r_{i1}^o = r_i^o - r_1^o$ is the range difference, $c^o$ indicates the unknown sound propagation speed and $n_{i1}$ is the zero-mean Gaussian TDOA noise.

Next, taking the time derivative of (1) results into the range rate as

$$\dot{r}_i^o = \frac{\left(\mathbf{u}^o - \mathbf{s}_i^o\right)^T \left(\dot{\mathbf{u}}^o - \dot{\mathbf{s}}_i^o\right)}{r_i^o}, \quad i = 1, 2, \ldots M. \tag{3}$$

and the FDOA measurement set is obtained, after divided with the signal carrier frequency, as

$$\dot{t}_{i1} = \frac{1}{c^o}\dot{r}_{i1}^o + \dot{n}_{i1}, \quad i = 2, 3, \ldots, M, \tag{4}$$

where $\dot{r}_{i1}^o = \dot{r}_i^o - \dot{r}_1^o$ is the rate of the range difference and $\dot{n}_{i1}$ is the zero-mean Gaussian FDOA noise.

Stacking the TDOA and FDOA measurements and fusing them into a $2(M-1)\times 1$ vector, we obtain

$$\boldsymbol{\alpha} \triangleq \left[\mathbf{t}^T, \dot{\mathbf{t}}^T\right]^T,$$

where

$$\mathbf{t} = \frac{1}{c^o}\left[r_{21}^o, \ldots, r_{M1}^o\right]^T + [n_{21}, \ldots, n_{M1}]^T = \frac{1}{c^o}\mathbf{r}^o + \mathbf{n},$$
$$\dot{\mathbf{t}} = \frac{1}{c^o}\left[\dot{r}_{21}^o, \ldots, \dot{r}_{M1}^o\right]^T + [\dot{n}_{21}, \ldots, \dot{n}_{M1}]^T = \frac{1}{c^o}\dot{\mathbf{r}}^o + \dot{\mathbf{n}}. \tag{5}$$

The measurement error vector is $\Delta\boldsymbol{\alpha} \triangleq \left[\mathbf{n}^T, \dot{\mathbf{n}}^T\right]^T$ and its covariance matrix is $\mathbf{Q}_{\boldsymbol{\alpha}} = E\left[\Delta\boldsymbol{\alpha}\Delta\boldsymbol{\alpha}^T\right]$. Note that the measurement noises $\Delta\boldsymbol{\alpha}$ are assumed to be independent of the SN parameter uncertainties $\Delta\boldsymbol{\beta}$.

In summary, the ultimate goal of this paper is inferring the location $\mathbf{u}^o$ and velocity $\dot{\mathbf{u}}^o$ of the source from the collected TDOA/FDOA measurements $\boldsymbol{\alpha}$ in presence of the unknown sound speed $c^o$ and the inaccurate SN parameters $\boldsymbol{\beta}$.

## III. CRAMÉR-RAO LOWER BOUND

The CRLB reveals the best expected performance for an unbiased estimator, which is very useful for studying the considered optimization problem and evaluating the proposed method. In this sectionïijŇ we first derive the CRLB for the estimator based on the TDOA/FDOA measurements in presence of the unknown sound speed $c^o$ and the inaccurate SN parameters $\boldsymbol{\beta}$, and then discuss the impact of an unknown sound propagation speed on the localization accuracy.

### A. HYBRID CRLB

Defining the parameter vector $\boldsymbol{\phi}^o \triangleq \left[(\boldsymbol{\theta}^o)^T, c^o, (\boldsymbol{\beta}^o)^T\right]^T$ with $\boldsymbol{\theta}^o \triangleq \left[(\mathbf{u}^o)^T, (\dot{\mathbf{u}}^o)^T\right]^T$, where $\boldsymbol{\theta}^o$ is a deterministic parameter, $c^o$ and $\boldsymbol{\beta}^o$ are random parameters. The resulting bound is the so-called Hybrid CRLB [33]–[35], which contains the classic CRLB and the Bayesian bound for deterministic parameters and random parameters, respectively. Recall that the TDOA/FDOA measurements vector $\boldsymbol{\alpha}$ and the SN parameter vector $\boldsymbol{\beta}$ are Gaussian distributed as $\boldsymbol{\alpha} \sim \mathcal{N}(\boldsymbol{\alpha}^o, \mathbf{Q}_{\boldsymbol{\alpha}})$ and $\boldsymbol{\beta} \sim \mathcal{N}(\boldsymbol{\beta}^o, \mathbf{Q}_{\boldsymbol{\beta}})$, respectively. Since $\boldsymbol{\alpha}$ is independent of $\boldsymbol{\beta}$, we can express the log-likelihood function for the joint probability density function of the measurements as

$$\begin{aligned}\ln\mathbb{P}(\mathbf{v};\boldsymbol{\phi}^o) &= \ln\mathbb{P}(\boldsymbol{\alpha}|\boldsymbol{\alpha}^o, \mathbf{Q}_{\boldsymbol{\alpha}}) + \ln\mathbb{P}(\boldsymbol{\beta}|\boldsymbol{\beta}^o, \mathbf{Q}_{\boldsymbol{\beta}})\\ &= C - \frac{1}{2}\left(\boldsymbol{\alpha}-\boldsymbol{\alpha}^o\right)^T\mathbf{Q}_{\boldsymbol{\alpha}}{}^{-1}\left(\boldsymbol{\alpha}-\boldsymbol{\alpha}^o\right)\\ &\quad - \frac{1}{2}\left(\boldsymbol{\beta}-\boldsymbol{\beta}^o\right)^T\mathbf{Q}_{\boldsymbol{\beta}}{}^{-1}\left(\boldsymbol{\beta}-\boldsymbol{\beta}^o\right),\end{aligned} \tag{6}$$

where the measurement vector $\mathbf{v} \triangleq \left[\boldsymbol{\alpha}^T, \boldsymbol{\beta}^T\right]^T$ is parameterized by $\boldsymbol{\phi}^o$ and $C$ collects the other constant terms. To obtain the CRLB, the Fisher information matrix (FIM) can be computed as [36]

$$\mathbf{I}(\boldsymbol{\phi}^o) = -E\left[\frac{\partial^2\ln\mathbb{P}(\mathbf{v};\boldsymbol{\phi}^o)}{\partial\boldsymbol{\phi}^o\partial(\boldsymbol{\phi}^o)^T}\right]. \tag{7}$$

For the convenience of expression, we rewrite $\mathbf{I}(\boldsymbol{\phi}^o)$ in submatrices as

$$\mathbf{I}(\boldsymbol{\phi}^o) = \begin{bmatrix}\mathbf{X}_{11} & \mathbf{X}_{12} & \mathbf{X}_{13}\\ \mathbf{X}_{12}^T & \mathbf{X}_{22} & \mathbf{X}_{23}\\ \mathbf{X}_{13}^T & \mathbf{X}_{23}^T & \mathbf{X}_{33}\end{bmatrix}, \tag{8}$$

where from (6), we have

$$\begin{aligned}\mathbf{X}_{11} &= -E\left[\frac{\partial^2\ln p(\mathbf{v};\boldsymbol{\phi})}{\partial\boldsymbol{\theta}^o\partial(\boldsymbol{\theta}^o)^T}\right] = \left(\frac{\partial\boldsymbol{\alpha}^o}{\partial\boldsymbol{\theta}^o}\right)^T\mathbf{Q}_{\boldsymbol{\alpha}}^{-1}\left(\frac{\partial\boldsymbol{\alpha}^o}{\partial\boldsymbol{\theta}^o}\right)\\ \mathbf{X}_{12} &= -E\left[\frac{\partial^2\ln p(\mathbf{v};\boldsymbol{\phi})}{\partial\boldsymbol{\theta}^o\partial c^o}\right] = \left(\frac{\partial\boldsymbol{\alpha}^o}{\partial\boldsymbol{\theta}^o}\right)^T\mathbf{Q}_{\boldsymbol{\alpha}}^{-1}\left(\frac{\partial\boldsymbol{\alpha}^o}{\partial c^o}\right)\\ \mathbf{X}_{13} &= -E\left[\frac{\partial^2\ln p(\mathbf{v};\boldsymbol{\phi})}{\partial\boldsymbol{\theta}^o\partial(\boldsymbol{\beta}^o)^T}\right] = \left(\frac{\partial\boldsymbol{\alpha}^o}{\partial\boldsymbol{\theta}^o}\right)^T\mathbf{Q}_{\boldsymbol{\alpha}}^{-1}\left(\frac{\partial\boldsymbol{\alpha}^o}{\partial\boldsymbol{\beta}^o}\right)\\ \mathbf{X}_{22} &= -E\left[\frac{\partial^2\ln p(\mathbf{v};\boldsymbol{\phi})}{\partial c^o\partial c^o}\right] = \left(\frac{\partial\boldsymbol{\alpha}^o}{\partial c^o}\right)^T\mathbf{Q}_{\boldsymbol{\alpha}}^{-1}\left(\frac{\partial\boldsymbol{\alpha}^o}{\partial c^o}\right)\\ \mathbf{X}_{23} &= -E\left[\frac{\partial^2\ln p(\mathbf{v};\boldsymbol{\phi})}{\partial c^o\partial(\boldsymbol{\beta}^o)^T}\right] = \left(\frac{\partial\boldsymbol{\alpha}^o}{\partial c^o}\right)^T\mathbf{Q}_{\boldsymbol{\alpha}}^{-1}\left(\frac{\partial\boldsymbol{\alpha}^o}{\partial\boldsymbol{\beta}^o}\right)\\ \mathbf{X}_{33} &= -E\left[\frac{\partial^2\ln p(\mathbf{v};\boldsymbol{\phi})}{\partial\boldsymbol{\beta}^o\partial(\boldsymbol{\beta}^o)^T}\right] = \left(\frac{\partial\boldsymbol{\alpha}^o}{\partial\boldsymbol{\beta}^o}\right)^T\mathbf{Q}_{\boldsymbol{\alpha}}^{-1}\left(\frac{\partial\boldsymbol{\alpha}^o}{\partial\boldsymbol{\beta}^o}\right) + \mathbf{Q}_{\boldsymbol{\beta}}^{-1}.\end{aligned} \tag{9}$$

Appendix A provides the further details for the partial derivatives in (9). Finally, the CRLB for the interested unknown parameters $\mathbf{u}^o$, $\dot{\mathbf{u}}^o$ and $c^o$ are obtained as following:

$$CRLB_{\mathbf{u}^o} = \sqrt{\sum_{k=1}^{3}\left[\mathbf{I}(\boldsymbol{\phi}^o)^{-1}\right]_{k,k}},$$
$$CRLB_{\dot{\mathbf{u}}^o} = \sqrt{\sum_{k=4}^{6}\left[\mathbf{I}(\boldsymbol{\phi}^o)^{-1}\right]_{k,k}}$$

and

$$CRLB_{c^o} = \sqrt{\left[\mathbf{I}(\boldsymbol{\phi}^o)^{-1}\right]_{7,7}}.$$

### B. THE IMPACT OF AN UNKNOWN SOUND PROPAGATION SPEED ON LOCALIZATION ACCURACY

In this subsection, we would like to study the impact of the sound propagation speed $c^o$ on the localization performance. When the SN parameters are subject to errors, we first consider the following two cases.

#### CASE 1: $C^O$ IS UNKNOWN

According to the block matrix inversion formula [36], we can formulate the inverse of the CRLB for the source parameter vector $\boldsymbol{\theta}^o$ from (8) as

$$CRLB_1(\boldsymbol{\theta}^o)^{-1} = \mathbf{X}_{11} - \begin{bmatrix} \mathbf{X}_{12} & \mathbf{X}_{13} \end{bmatrix} \begin{bmatrix} \mathbf{X}_{22} & \mathbf{X}_{23} \\ \mathbf{X}_{23}^T & \mathbf{X}_{33} \end{bmatrix}^{-1} \begin{bmatrix} \mathbf{X}_{12}^T \\ \mathbf{X}_{13}^T \end{bmatrix} \tag{10}$$

Here, note that $CRLB_1(\boldsymbol{\theta}^o)$ reduces to $\mathbf{X}_{11}^{-1}$ when there is no SN parameter error and a known sound propagation speed $c^o$ is considered. This implies that the trace of $\mathbf{X}_{11}^{-1}$ is the lower bound for the mean squared error (MSE) of any unbiased estimator for $\boldsymbol{\theta}^o$ that uses the TDOA/FDOA measurements, i.e, with all the necessary and exactly accurate knowledge. On the other hand, we can readily observe that the second term in (10) collects the impacts of the SN parameter errors and the unknown sound propagation speed on the CRLB of $\boldsymbol{\theta}^o$, which are apparently not incurred in an additive manner.

For the convenience of comparing with the counterpart of *Case 2* in the following, we reformulate (10) into a different form

$$\begin{aligned} CRLB_1(\boldsymbol{\theta}^o)^{-1} &= \left(\frac{\partial \boldsymbol{\alpha}^o}{\partial \boldsymbol{\theta}^o}\right)^T \mathbf{Q}_1^{-1} \left(\frac{\partial \boldsymbol{\alpha}^o}{\partial \boldsymbol{\theta}^o}\right) \\ &\quad - \left(\frac{\partial \boldsymbol{\alpha}^o}{\partial \boldsymbol{\theta}^o}\right)^T \mathbf{Q}_1^{-1} \left(\frac{\partial \boldsymbol{\alpha}^o}{\partial \boldsymbol{\beta}^o}\right) \boldsymbol{\Gamma}_1 \left(\frac{\partial \boldsymbol{\alpha}^o}{\partial \boldsymbol{\beta}^o}\right)^T \mathbf{Q}_1^{-1} \left(\frac{\partial \boldsymbol{\alpha}^o}{\partial \boldsymbol{\theta}^o}\right), \end{aligned} \tag{11}$$

where

$$\begin{aligned} \mathbf{Q}_1^{-1} &= \mathbf{Q}_{\boldsymbol{\alpha}}^{-1} \\ &\quad - \mathbf{Q}_{\boldsymbol{\alpha}}^{-1} \left(\frac{\partial \boldsymbol{\alpha}^o}{\partial c^o}\right) \left(\left(\frac{\partial \boldsymbol{\alpha}^o}{\partial c^o}\right)^T \mathbf{Q}_{\boldsymbol{\alpha}}^{-1} \left(\frac{\partial \boldsymbol{\alpha}^o}{\partial c^o}\right)\right)^{-1} \\ &\quad \times \left(\frac{\partial \boldsymbol{\alpha}^o}{\partial c^o}\right)^T \mathbf{Q}_{\boldsymbol{\alpha}}^{-1}, \end{aligned} \tag{12}$$

and

$$\Gamma_1 = \left(\left(\frac{\partial \boldsymbol{\alpha}^o}{\partial \boldsymbol{\beta}^o}\right)^T \mathbf{Q}_1^{-1} \left(\frac{\partial \boldsymbol{\alpha}^o}{\partial \boldsymbol{\beta}^o}\right) + \mathbf{Q}_{\boldsymbol{\beta}}^{-1}\right)^{-1}. \tag{13}$$

Please refer to Appendix B for the detailed derivations.

### CASE 2: $C^O$ IS KNOWN

In this scenario, since $c^o$ is known, there is no partial derivative over $c^o$. Similar to the steps described in Section III-A, we can easily obtain the inverse of the CRLB for estimating the source parameter vector $\boldsymbol{\theta}^o$ in this case as

$$\begin{aligned} CRLB_2(\boldsymbol{\theta}^o)^{-1} &= \mathbf{X}_{11} - \mathbf{X}_{13}\mathbf{X}_{33}^{-1}\mathbf{X}_{13}^T \\ &= \left(\frac{\partial \boldsymbol{\alpha}^o}{\partial \boldsymbol{\theta}^o}\right)^T \mathbf{Q}_{\boldsymbol{\alpha}}^{-1} \left(\frac{\partial \boldsymbol{\alpha}^o}{\partial \boldsymbol{\theta}^o}\right) \\ &\quad - \left(\frac{\partial \boldsymbol{\alpha}^o}{\partial \boldsymbol{\theta}^o}\right)^T \mathbf{Q}_{\boldsymbol{\alpha}}^{-1} \left(\frac{\partial \boldsymbol{\alpha}^o}{\partial \boldsymbol{\beta}^o}\right) \boldsymbol{\Gamma}_2 \left(\frac{\partial \boldsymbol{\alpha}^o}{\partial \boldsymbol{\beta}^o}\right)^T \\ &\quad \times \mathbf{Q}_{\boldsymbol{\alpha}}^{-1} \left(\frac{\partial \boldsymbol{\alpha}^o}{\partial \boldsymbol{\theta}^o}\right), \end{aligned} \tag{14}$$

where

$$\Gamma_2 = \left(\left(\frac{\partial \boldsymbol{\alpha}^o}{\partial \boldsymbol{\beta}^o}\right)^T \mathbf{Q}_{\boldsymbol{\alpha}}^{-1} \left(\frac{\partial \boldsymbol{\alpha}^o}{\partial \boldsymbol{\beta}^o}\right) + \mathbf{Q}_{\boldsymbol{\beta}}^{-1}\right)^{-1}.$$

### DISCUSSIONS

Obviously, from (11) and (14), the covariance matrices $\mathbf{Q}_1$ and $\mathbf{Q}_{\boldsymbol{\alpha}}$ differentiate the CRLBs with a known or unknown $c^o$, and hence our breakthrough point lies in (12). Let us denote $\mathbf{Q}_{\boldsymbol{\alpha}}^{-1/2}$ as the Cholesky decomposition of $\mathbf{Q}_{\boldsymbol{\alpha}}^{-1}$, i.e. $\mathbf{Q}_{\boldsymbol{\alpha}}^{-1} = \mathbf{Q}_{\boldsymbol{\alpha}}^{-1/2}\mathbf{Q}_{\boldsymbol{\alpha}}^{-1/2}$. Then, the projection matrix of the subspace spanned by the columns of $\left(\frac{\partial \boldsymbol{\alpha}^o}{\partial c^o}\right)^T \mathbf{Q}_{\boldsymbol{\alpha}}^{-1/2}$ can be written as

$$\begin{aligned} \mathbf{P}_1 &= \mathbf{Q}_{\boldsymbol{\alpha}}^{-1/2} \left(\frac{\partial \boldsymbol{\alpha}^o}{\partial c^o}\right) \left(\left(\frac{\partial \boldsymbol{\alpha}^o}{\partial c^o}\right)^T \mathbf{Q}_{\boldsymbol{\alpha}}^{-1} \left(\frac{\partial \boldsymbol{\alpha}^o}{\partial c^o}\right)\right)^{-1} \\ &\quad \times \left(\frac{\partial \boldsymbol{\alpha}^o}{\partial c^o}\right)^T \mathbf{Q}_{\boldsymbol{\alpha}}^{-1/2}. \end{aligned} \tag{15}$$

Accordingly, the orthogonal projection matrix of $\mathbf{P}_1$ is $\mathbf{P}_1^{\perp} = \mathbf{I} - \mathbf{P}_1$. Then, (12) can readily be rewritten into

$$\mathbf{Q}_1^{-1} = \mathbf{Q}_{\boldsymbol{\alpha}}^{-1/2}\mathbf{P}_1^{\perp}\mathbf{Q}_{\boldsymbol{\alpha}}^{-1/2}, \tag{16}$$

where we observe that the unknown sound propagation speed affects the measurement noise covariance matrix by projecting its square root inverse ($\mathbf{Q}_{\boldsymbol{\alpha}}^{-1/2}$) onto the orthogonal subspace spanned by the columns of $\left(\frac{\partial \boldsymbol{\alpha}^o}{\partial c^o}\right)^T \mathbf{Q}_{\boldsymbol{\alpha}}^{-1/2}$. In a nutshell, the impact of an unknown $c^o$ manifests itself on the covariance matrix in a projecting rather than additive manner.

#### *Numerical Study*

To be more explicit, we hereby show the impact of an unknown $c^o$ on the CRLB via the numerical study. We evaluate the corresponding CRLBs in both cases under different SN parameter error variances and measurement noise variances, respectively. In the simulations, there are 10 SNs and their true position and velocities are tabulated in Table 1. The source to be located is at $[200,800,200]^T$ with instantaneous velocity $\dot{\mathbf{u}}^o = [-2, 1.5, 1]^T$ m/s. The noisy SN parameters are generated by adding the true values zero-mean white Gaussian noise with covariance matrix $\mathbf{Q}_{\boldsymbol{\beta}} = \sigma_s^2 diag\,([\mathbf{b}, 0.5\mathbf{b}])$, where $\mathbf{b} = [1, 20, 10, 30, 20, 3, 2, 10, 1, 2] \otimes \mathbf{1}_3^T$. The TDOA and FDOA measurements are generated with covariance matrix $\mathbf{Q}_\alpha = \sigma_d^2/c^{o2} blkdiag\,(\mathbf{R}, 0.1\mathbf{R})$, where $\mathbf{R}$ is a $(M-1)\times(M-1)$ matrix with unity on its diagonal and 0.5 in all other elements.

Fig. 2(a) illustrates the position and velocity estimation accuracy (under both with and without sound propagation speed knowledge) versus SN parameter error variance. The measurement noise parameter $\sigma_d$ is set to be 1 m. We observe that the estimation accuracy of the case without sound propagation speed knowledge is worse than the case with sound propagation speed knowledge. For example, at the SN parameter error variance $\sigma_s^2 = 1$, i.e. $10\log(\sigma_s^2) = 0$, the position performance gap between both cases is 1.07 dB and that for the velocity is 0.96 dB. What's more, when the SN parameter error decreases, the performance gap between both cases increases slightly. Fig. 2(b) shows the results versus measurement noise variance. The SN parameter error $\sigma_s$ is fixed to be 1 m. The increases in CRLB is also observed assuming the sound propagation speed as an unknown parameter. Therefore, we conclude that the sound propagation speed should be well considered to obtain a more accurate source parameter estimates.

**TABLE 1. True positions (in meters) and velocities (in meters/second) of SNs.**

| SN no. $i$ | $x_i^o$ | $y_i^o$ | $z_i^o$ | $\dot{x}_i^o$ | $\dot{y}_i^o$ | $\dot{z}_i^o$ |
|---|---|---|---|---|---|---|
| 1 | 0 | 1000 | 0 | 3 | -2 | 2 |
| 2 | 0 | 0 | 0 | -3 | 1 | 2 |
| 3 | 0 | 0 | 1000 | 1 | -2 | 1 |
| 4 | 0 | 1000 | 1000 | 1 | 2 | 3 |
| 5 | 1000 | 0 | 0 | -2 | 1 | 1 |
| 6 | 1000 | 1000 | 0 | 2 | -1 | 1 |
| 7 | 1000 | 0 | 1000 | 1.2 | -1.5 | 1.5 |
| 8 | 1000 | 1000 | 1000 | -1.5 | 1.2 | -1.2 |
| 9 | 500 | 500 | 1000 | 1.3 | 1.3 | 1.3 |
| 10 | 500 | 500 | 0 | 2.5 | 2.5 | 2.5 |

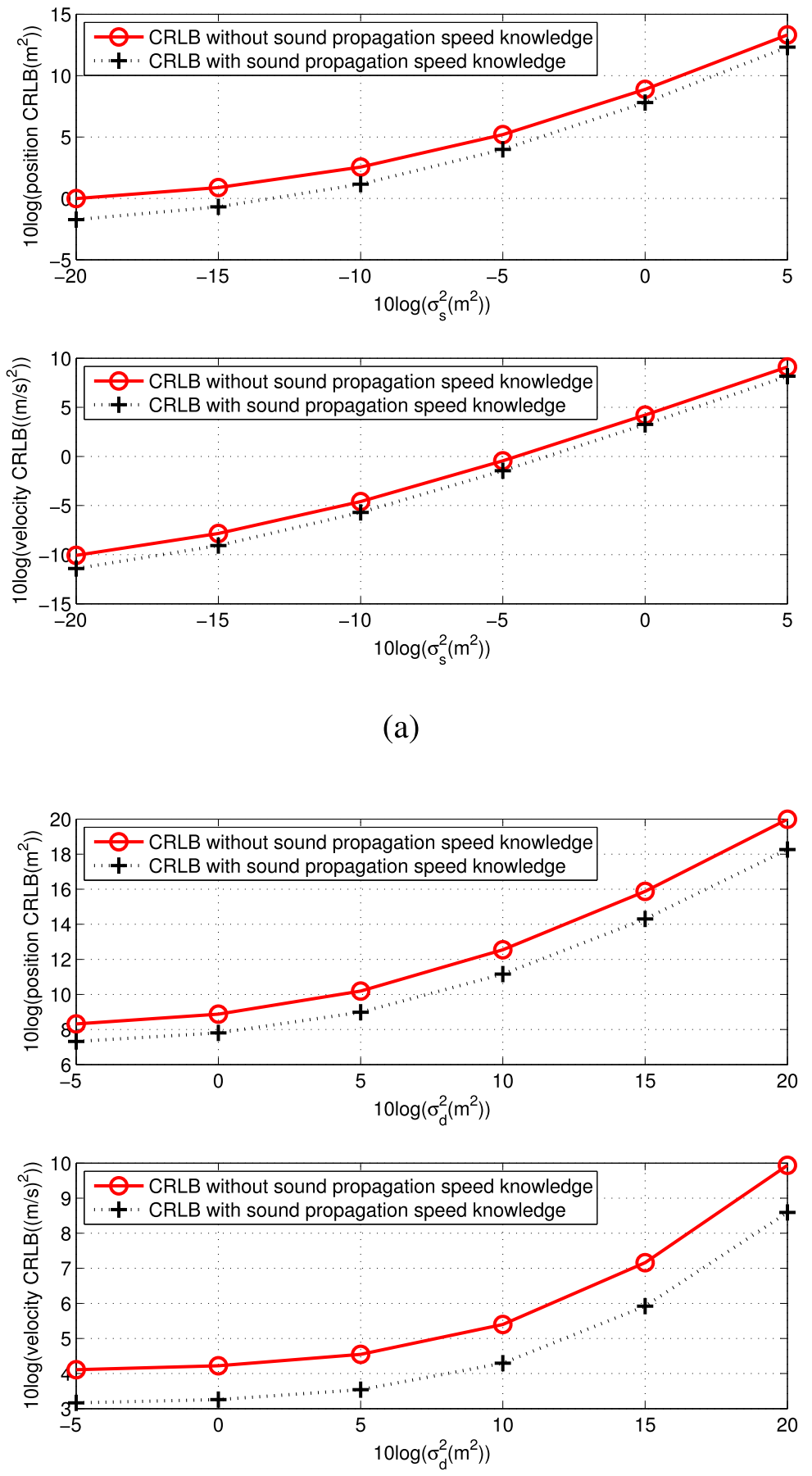

**FIGURE 2. Comparisons of CRLBs with and without sound propagation speed knowledge: (a) CRLBs versus SN parameter error; (b) CRLBs versus measurement noise.**

## IV. PROPOSED UNDERWATER LOCALIZATION APPROACH

In this section, we would like to introduce our solution to the underwater localization problem formulated in Section II. Again, note that the previous literature either consider a known sound propagation speed [13], [37] or assume the SN parameters to be exactly known [32], both of which are practically very difficult to guarantee and hence rather unrealistic.

In case of an unknown sound propagation speed and inaccurate SN parameters, our proposed approach includes two stages while dealing with the considered underwater localization problem. The first stage linearizes the measurement equations by introducing nuisance variables, where we can obtain an initial solution. Then, the second stage exploits the relationship between the source parameters and the nuisance variables such that we can refine the solution obtained in the first step. To be specific, we elaborate the aforementioned two stages as follows.

### A. FIRST STAGE

For the use of the TDOA and FDOA measurements, we have to fuse these two kinds of measurements by linearizing their equations such that we can obtain our linear data model and then initially estimate the location $\mathbf{u}^o$ and velocity $\dot{\mathbf{u}}^o$ of the source.

#### 1) TDOA MEASUREMENTS

After rewriting the range difference as $r_{i1}^o + r_1^o = r_i^o$, squaring both its sides and plugging (1) into the terms of $r_1^{o2}$ and $r_i^{o2}$, we obtain a set of TDOA equations as

$$r_{i1}^{o\,2} + 2r_{i1}^o r_1^o = \mathbf{s}_i^{oT}\mathbf{s}_i^o - \mathbf{s}_1^{oT}\mathbf{s}_1^o - 2\big(\mathbf{s}_i^o - \mathbf{s}_1^o\big)^T\mathbf{u}^o, \quad (17)$$

for $i = 2, 3, \ldots, M$.

Then, we take the noise terms into account by substituting $r_{i1}^o = c^o t_{i1} - c^o n_{i1}$ and $\mathbf{s}_i^o = \mathbf{s}_i - \Delta\mathbf{s}_i$ into (17), which leads to

$$\begin{aligned} &2c^o r_i^o n_{i1} + 2\big(\mathbf{u}^o - \mathbf{s}_i\big)^T\Delta\mathbf{s}_i - 2\big(\mathbf{u}^o - \mathbf{s}_1\big)^T\Delta\mathbf{s}_1 \\ &\quad\approx R_1 - R_i + 2(\mathbf{s}_i - \mathbf{s}_1)^T\mathbf{u}^o + 2t_{i1}c^o r_1^o + t_{i1}^2 c^{o2}, \quad (18) \end{aligned}$$

where $R_i \triangleq \mathbf{s}_i^T\mathbf{s}_i$ and all the second order error terms are relatively insignificant hence ignored. Here, note that the term $r_1^o$ is still dependent on the true value $\mathbf{s}_1^o$. Thus, we apply the first order Taylor approximation to $r_1^o$ around the noisy sensor position $\mathbf{s}_1$ as

$$r_1^o = \big\|\mathbf{u}^o - \mathbf{s}_1^o\big\| \approx \hat{r}_1^o + \boldsymbol{\rho}_{\mathbf{u}^o,\mathbf{s}_1}^T\Delta\mathbf{s}_1, \quad (19)$$

where $\hat{r}_1^o \triangleq \|\mathbf{u}^o - \mathbf{s}_1\|$ is the distance between the source node and the nominal known location of the first SN, and $\boldsymbol{\rho}_{\mathbf{a},\mathbf{b}} \triangleq \mathbf{a} - \mathbf{b}\big/\|\mathbf{a} - \mathbf{b}\|$ represents the unit vector from $\mathbf{b}$ to $\mathbf{a}$. Substituting (19) into (18) and gathering all the noise terms

into $\varepsilon_{t,i}$ at the left side, we obtain

$$\begin{aligned}&2c^o r_i^o n_{i1} + 2(\mathbf{u}^o - \mathbf{s}_i)^T \Delta\mathbf{s}_i - 2\mathbf{d}_{i1}^T \Delta\mathbf{s}_1\\ &\quad\approx R_1 - R_i + 2(\mathbf{s}_i - \mathbf{s}_1)^T \mathbf{u}^o + 2t_{i1}\eta_1 + t_{i1}^2\eta_2,\\ &\quad\Rightarrow \varepsilon_{t,i} \approx R_1 - R_i + 2(\mathbf{s}_i - \mathbf{s}_1)^T \mathbf{u}^o + 2t_{i1}\eta_1 + t_{i1}^2\eta_2, \quad (20)\end{aligned}$$

where $\eta_1 \triangleq c^o\hat{r}_1^o$, $\eta_2 \triangleq c^{o2}$ and $\mathbf{d}_{i1} \triangleq \mathbf{u}^o - \mathbf{s}_1 + c^o t_{i1}\boldsymbol{\rho}_{\mathbf{u}^o,\mathbf{s}_1}$. Obviously, (20) is already a pseudo-linear equation w.r.t. unknowns $\mathbf{u}^o$, $\eta_1$ and $\eta_2$, where $\eta_1$ and $\eta_2$ are the nuisance variables that we introduced to facilitate the derivations.

### 2) FDOA MEASUREMENTS

Taking the time derivative of (17) results in a set of FDOA equations as

$$\begin{aligned}&r_{i1}^o\dot{r}_{i1}^o + r_{i1}^o\dot{r}_1^o + \dot{r}_{i1}^o r_1^o\\ &\quad= \dot{\mathbf{s}}_i^{oT}\mathbf{s}_i^o - \dot{\mathbf{s}}_1^{oT}\mathbf{s}_1^o - (\dot{\mathbf{s}}_i^o - \dot{\mathbf{s}}_1^o)^T\mathbf{u}^o - (\mathbf{s}_i^o - \mathbf{s}_1^o)^T\dot{\mathbf{u}}^o, \quad (21)\end{aligned}$$

for $i = 2, 3, \ldots, M$. Substituting $r_{i1}^o = c^o t_{i1} - c^o n_{i1}$, $\dot{r}_{i1}^o = c^o\dot{t}_{i1} - c^o\dot{n}_{i1}$, $\mathbf{s}_i^o = \mathbf{s}_i - \Delta\mathbf{s}_i$ and $\dot{\mathbf{s}}_i^o = \dot{\mathbf{s}}_i - \Delta\dot{\mathbf{s}}_i$ into (21) and ignoring the second-order error terms, we obtain

$$\begin{aligned}&c^o r_i^o\dot{n}_{i1} + c^o\dot{r}_i^o n_{i1} + (\dot{\mathbf{u}}^o - \dot{\mathbf{s}}_i)^T\Delta\mathbf{s}_i - (\dot{\mathbf{u}}^o - \dot{\mathbf{s}}_1)^T\Delta\mathbf{s}_1\\ &\qquad + (\mathbf{u}^o - \mathbf{s}_i)^T\Delta\dot{\mathbf{s}}_i - (\mathbf{u}^o - \mathbf{s}_1)^T\Delta\dot{\mathbf{s}}_1\\ &\qquad\approx \dot{R}_1 - \dot{R}_i + (\dot{\mathbf{s}}_i - \dot{\mathbf{s}}_1)^T\mathbf{u}^o + (\mathbf{s}_i - \mathbf{s}_1)^T\dot{\mathbf{u}}^o\\ &\qquad + c^{o2}t_{i1}\dot{t}_{i1} + c^o t_{i1}\dot{r}_1^o + \dot{t}_{i1}c^o r_1^o, \quad (22)\end{aligned}$$

where $\dot{R}_i \triangleq \dot{\mathbf{s}}_i^T\mathbf{s}_i$. In addition to $r_1^o$, $\dot{r}_1^o$ depends on the true values $\mathbf{s}_1^o$ and $\dot{\mathbf{s}}_1^o$. Similarly, we apply the Taylor series approximation around $\mathbf{s}_1$ and we have

$$\dot{r}_1^o \approx \hat{\dot{r}}_1^o + \boldsymbol{\lambda}_{\mathbf{u}^o,\mathbf{s}_1}^T\Delta\mathbf{s}_1 + \boldsymbol{\rho}_{\mathbf{u}^o,\mathbf{s}_1}^T\Delta\dot{\mathbf{s}}_1, \quad (23)$$

where $\hat{\dot{r}}_1^o \triangleq (\mathbf{u}^o - \mathbf{s}_1)^T(\dot{\mathbf{u}}^o - \dot{\mathbf{s}}_1)/\hat{r}_1^o$ and $\boldsymbol{\lambda}_{\mathbf{u}^o,\mathbf{s}_1} \triangleq \frac{\dot{\mathbf{u}}^o - \dot{\mathbf{s}}_1}{\hat{r}_1^{o2}} - \frac{\hat{\dot{r}}_1^o}{\hat{r}_1^o}\boldsymbol{\rho}_{\mathbf{u}^o,\mathbf{s}_1}$. Then, we plug (23) into (22) and again collect the noise terms into $\varepsilon_{\dot{t},i}$ at the left side, leading to our linear equation for the FDOA measurements as

$$\begin{aligned}&c^o r_i^o\dot{n}_{i1} + c^o\dot{r}_i^o n_{i1} + (\dot{\mathbf{u}}^o - \dot{\mathbf{s}}_i)^T\Delta\mathbf{s}_i\\ &\quad - \dot{\mathbf{d}}_{i1}^T\Delta\mathbf{s}_1 + (\mathbf{u}^o - \mathbf{s}_i)^T\Delta\dot{\mathbf{s}}_i - \mathbf{d}_{i1}{}^T\Delta\dot{\mathbf{s}}_1\\ &\quad\approx \dot{R}_1 - \dot{R}_i + (\dot{\mathbf{s}}_i - \dot{\mathbf{s}}_1)^T\mathbf{u}^o + (\mathbf{s}_i - \mathbf{s}_1)^T\dot{\mathbf{u}}^o\\ &\quad + \dot{t}_{i1}\eta_1 + t_{i1}\dot{t}_{i1}\eta_2 + t_{i1}\dot{\eta}_1,\\ &\quad\Rightarrow \varepsilon_{\dot{t},i} \approx \dot{R}_1 - \dot{R}_i + (\dot{\mathbf{s}}_i - \dot{\mathbf{s}}_1)^T\mathbf{u}^o + (\mathbf{s}_i - \mathbf{s}_1)^T\dot{\mathbf{u}}^o\\ &\quad + \dot{t}_{i1}\eta_1 + t_{i1}\dot{t}_{i1}\eta_2 + t_{i1}\dot{\eta}_1, \quad (24)\end{aligned}$$

where $\dot{\eta}_1 \triangleq c^o\hat{\dot{r}}_1^o$ and $\dot{\mathbf{d}}_{i1} \triangleq \dot{\mathbf{u}}^o - \dot{\mathbf{s}}_1 + c^o t_{i1}\boldsymbol{\lambda}_{\mathbf{u}^o,\mathbf{s}_1} + c^o\dot{t}_{i1}\boldsymbol{\rho}_{\mathbf{u}^o,\mathbf{s}_1}$. Finally, (24) also becomes linear w.r.t. unknowns $\mathbf{u}^o$, $\dot{\mathbf{u}}^o$, $\eta_1$, $\eta_2$ and $\dot{\eta}_1$, where $\dot{\eta}_1$ is another introduced nuisance parameter.

### 3) MEASUREMENT FUSION AND DATA MODELS

Let us collect the source parameters and the three nuisance parameters into a single parameter vector as $\boldsymbol{\varphi}_1^o \triangleq [\mathbf{u}^o, \dot{\mathbf{u}}^o, \eta_1, \eta_2, \dot{\eta}_1]^T$, on which we can base to fomulate our linear data model.

For the TDOA measurements, stacking (20) for $i = 2, 3, \ldots, M$ into vectors results our linear data model with $\boldsymbol{\varphi}_1^o$ as

$$\boldsymbol{\varepsilon}_t = \mathbf{h}_t - \mathbf{G}_t\boldsymbol{\varphi}_1^o, \quad (25)$$

where

$$\mathbf{h}_t \triangleq \begin{bmatrix} R_1 - R_2\\ R_1 - R_3\\ \vdots\\ R_1 - R_M\end{bmatrix},$$

$$\mathbf{G}_t \triangleq -\begin{bmatrix} 2(\mathbf{s}_2 - \mathbf{s}_1)^T & 0 & 2t_{21} & t_{21}^2 & 0\\ 2(\mathbf{s}_3 - \mathbf{s}_1)^T & 0 & 2t_{31} & t_{31}^2 & 0\\ \vdots & \vdots & \vdots & \vdots & \vdots\\ 2(\mathbf{s}_M - \mathbf{s}_1)^T & 0 & 2t_{M1} & t_{M1}^2 & 0\end{bmatrix}.$$

The noise vector $\boldsymbol{\varepsilon}_t$ is defined as

$$\boldsymbol{\varepsilon}_t \triangleq \begin{bmatrix}\mathbf{B} & \mathbf{0}_{(M-1)\times(M-1)}\end{bmatrix}\Delta\boldsymbol{\alpha} + \begin{bmatrix}\mathbf{D} & \mathbf{0}_{(M-1)\times 3M}\end{bmatrix}\Delta\boldsymbol{\beta}, \quad (26)$$

where $\mathbf{B} = \mathrm{diag}\left(2\left[c^o r_2^o, c^o r_3^o, \cdots, c^o r_M^o\right]\right)$ and $\mathbf{D}$ is an $(M-1)\times 3M$ matrix whose $(i-1)$th row, $i = 2, 3, \ldots, M$, is $2\left[-\mathbf{d}_{i1}^T \;\; \mathbf{0}_{1\times 3(i-2)} \;\; (\mathbf{u}^o - \mathbf{s}_i)^T \;\; \mathbf{0}_{1\times 3(M-i)}\right]$.

For the FDOA measurements, similarly stacking (24) for $i = 2, 3, \ldots, M$ into vectors, we obtain our linear data model with $\boldsymbol{\varphi}_1^o$ as

$$\boldsymbol{\varepsilon}_{\dot{t}} = \mathbf{h}_{\dot{t}} - \mathbf{G}_{\dot{t}}\boldsymbol{\varphi}_1^o, \quad (27)$$

where

$$\mathbf{h}_{\dot{t}} \triangleq 2\begin{bmatrix}\dot{R}_1 - \dot{R}_2\\ \dot{R}_1 - \dot{R}_3\\ \vdots\\ \dot{R}_1 - \dot{R}_M\end{bmatrix},$$

$$\mathbf{G}_{\dot{t}} \triangleq -2\begin{bmatrix}(\dot{\mathbf{s}}_2 - \dot{\mathbf{s}}_1)^T & (\mathbf{s}_2 - \mathbf{s}_1)^T & \dot{t}_{21} & t_{21}\dot{t}_{21} & t_{21}\\ (\dot{\mathbf{s}}_3 - \dot{\mathbf{s}}_1)^T & (\mathbf{s}_3 - \mathbf{s}_1)^T & \dot{t}_{31} & t_{31}\dot{t}_{31} & t_{31}\\ \vdots & \vdots & \vdots & \vdots & \vdots\\ (\dot{\mathbf{s}}_M - \dot{\mathbf{s}}_1)^T & (\mathbf{s}_M - \mathbf{s}_1)^T & \dot{t}_{M1} & t_{M1}\dot{t}_{M1} & t_{M1}\end{bmatrix}.$$

The noise vector $\boldsymbol{\varepsilon}_{\dot{t}}$ is defined as

$$\boldsymbol{\varepsilon}_{\dot{t}} = \begin{bmatrix}\dot{\mathbf{B}} & \mathbf{B}\end{bmatrix}\Delta\boldsymbol{\alpha} + \begin{bmatrix}\dot{\mathbf{D}} & \mathbf{D}\end{bmatrix}\Delta\boldsymbol{\beta}, \quad (28)$$

where $\dot{\mathbf{B}} \triangleq \mathrm{diag}\left(2\left[c^o\dot{r}_2^o, c^o\dot{r}_3^o, \cdots, c^o\dot{r}_M^o\right]\right)$ and $\dot{\mathbf{D}}$ is an $(M-1)\times 3M$ matrix whose $(i-1)$th row, $i = 2, 3, \ldots, M$, is $2\left[-\dot{\mathbf{d}}_{i1}^T \;\; \mathbf{0}_{1\times 3(i-2)} \;\; (\dot{\mathbf{u}}^o - \dot{\mathbf{s}}_i)^T \;\; \mathbf{0}_{1\times 3(M-i)}\right]$.

Finally, we can easily fuse this two kinds of measurements by combining (25) and (27) as

$$\boldsymbol{\varepsilon}_1 = \mathbf{h}_1 - \mathbf{G}_1\boldsymbol{\varphi}_1^o, \quad (29)$$

where $\boldsymbol{\varepsilon}_1 \triangleq \left[\boldsymbol{\varepsilon}_t^T \; \boldsymbol{\varepsilon}_{\dot{t}}^T\right]^T$, $\mathbf{h}_1 \triangleq \left[\mathbf{h}_t^T \; \mathbf{h}_{\dot{t}}^T\right]^T$ and $\mathbf{G}_1 \triangleq \left[\mathbf{G}_t^T \; \mathbf{G}_{\dot{t}}^T\right]^T$.

#### 4) INITIAL SOLUTION

Minimizing the weighted square norm of $\boldsymbol{\varepsilon}_1$ yields the WLS solution

$$\boldsymbol{\varphi}_1 = \left(\mathbf{G}_1^T\mathbf{W}_1\mathbf{G}_1\right)^{-1}\mathbf{G}_1^T\mathbf{W}_1\mathbf{h}_1, \tag{30}$$

where $\mathbf{W}_1$ is the weighting matrix defined as $\mathbf{W}_1 = E\left(\boldsymbol{\varepsilon}_1\boldsymbol{\varepsilon}_1^T\right)^{-1}$. From (26) and (28), we have

$$\boldsymbol{\varepsilon}_1 = \mathbf{B}_1\Delta\boldsymbol{\alpha} + \mathbf{D}_1\Delta\boldsymbol{\beta}, \tag{31}$$

where

$$\mathbf{B}_1 = \begin{bmatrix} \mathbf{B} & \mathbf{0}_{(M-1)\times(M-1)} \\ \dot{\mathbf{B}} & \mathbf{B} \end{bmatrix}, \quad \mathbf{D}_1 = \begin{bmatrix} \mathbf{D} & \mathbf{0}_{(M-1)\times 3M} \\ \dot{\mathbf{D}} & \mathbf{D} \end{bmatrix}.$$

Hence, $\mathbf{W}_1$ can be approximately calculated as

$$\mathbf{W}_1 = \left(\mathbf{B}_1\mathbf{Q}_{\alpha}\mathbf{B}_1^T + \mathbf{D}_1\mathbf{Q}_{\beta}\mathbf{D}_1^T\right)^{-1}, \tag{32}$$

where the assumption of the independent relationship between the measurement noise $\Delta\boldsymbol{\alpha}$ and the SN's position uncertainty $\Delta\boldsymbol{\beta}$ is used. We also would like to point out that $\mathbf{D}_1$ contains the noisy measurements. However, the noise is ignored when taking the expectation by assuming the noise is of small value.

We shall evaluate the covariance matrix of the estimate $\boldsymbol{\varphi}_1$ by first representing the estimation error as

$$\begin{aligned}\Delta\boldsymbol{\varphi}_1 = \boldsymbol{\varphi}_1 - \boldsymbol{\varphi}_1^o &= \left(\mathbf{G}_1^T\mathbf{W}_1\mathbf{G}_1\right)^{-1}\mathbf{G}_1^T\mathbf{W}_1\left(\mathbf{h}_1 - \mathbf{G}_1\boldsymbol{\varphi}_1^o\right) \\ &= \left(\mathbf{G}_1^T\mathbf{W}_1\mathbf{G}_1\right)^{-1}\mathbf{G}_1^T\mathbf{W}_1\boldsymbol{\varepsilon}_1. \end{aligned} \tag{33}$$

When the noise in $\mathbf{G}_1$ is small enough to be ignored, the covariance matrix of $\boldsymbol{\varphi}_1$ can be approximated by

$$cov\left(\boldsymbol{\varphi}_1\right) = \left(\mathbf{G}_1^T\mathbf{W}_1\mathbf{G}_1\right)^{-1}. \tag{34}$$

In fact, there exist two practical issues while computing (30). First, the knowledge of the error covariance matrices $\mathbf{Q}_{\alpha}$ and $\mathbf{Q}_{\beta}$. In practical, they are usually obtained by pre-calibration. The statistical characteristics of the errors can be extracted from the field measurements. Second, the calculation of the weighting matrix $\mathbf{W}_1$ requires the true values of the parameter to be estimated. To cope with this, $\mathbf{W}_1 = \mathbf{Q}_{\alpha}^{-1}$ is first used to calculate (30), then substituting the initial estimate into (32) to obtain an improved version of $\mathbf{W}_1$ which is sequentially applied to obtain a better estimate of $\boldsymbol{\varphi}_1$. As a result, there exists an iteration between the updating of $\mathbf{W}_1$ and $\boldsymbol{\varphi}_1$. However, as observed in our implementation, iterating the solution computation only one time is sufficient to yield an accurate result. Additional iteration does not affect the solution accuracy significantly.

Unfortunately, in some cases, the measurement error covariance matrices $\mathbf{Q}_{\alpha}$ and $\mathbf{Q}_{\beta}$ might be unavailable due to the pre-calibration cost. Thus, we cannot use $\mathbf{W}_1 = \mathbf{Q}_{\alpha}^{-1}$ to calculate (30). In this situation, we actually have two choices to cope with the weighting matrix $\mathbf{W}_1$:

1) At the beginning, $\mathbf{W}_1 = \mathbf{I}$ is first used to calculate (30), then replacing $\mathbf{Q}_{\alpha}$ and $\mathbf{Q}_{\beta}$ with the identity matrix and computing $\mathbf{W}_1$ from (32) to poduce a better solution. Since $\mathbf{B}_1$ and $\mathbf{D}_1$ are used in updating $\mathbf{W}_1$, the structure information of $\mathbf{W}_1$ is retained although we have no information about the error covariance matrices. As a result, the localization performance degradation can be alleviated.
2) Using $\mathbf{W}_1 = \mathbf{I}$ to calculate (30) directly and the $\mathbf{W}_1$ updating procedure is omitted. In this manner, the WLS estimation in the first stage is reduced to the LS estimation. More importantly, besides the error covariance information, the structure information of the weighting matrix $\mathbf{W}_1$ is also ignored. Thus, this kind of handing $\mathbf{W}_1$ would introduce larger estimation error compared with the one described in 1).

The impact of the treatments of the weighting matrix $\mathbf{W}_1$ with/without error covariance matrices on the performance of the proposed solution will be compared and discussed in the simulation section.

### B. SECOND STAGE

In this stage, we shall refine the estimate obtained in the first stage by exploiting the relationship between the parameters in $\boldsymbol{\varphi}_1^o$. When the measurement noise and sensor uncertainty are small enough to be ignored, the estimate $\boldsymbol{\varphi}_1$ is unbiased according to the WLS theroy [36]. Hence, $\boldsymbol{\varphi}_1$ can be considered as a random vector with mean $\boldsymbol{\varphi}_1^o$ and covariance matrix $cov\left(\boldsymbol{\varphi}_1\right)$. The second-order error terms are ignored in the following derivation.

The first three elements in $\boldsymbol{\varphi}_1$, thus can be expressed as $\boldsymbol{\varphi}_1(1:3) = \mathbf{u}^o + \Delta\boldsymbol{\varphi}_1(1:3)$. Subtracting both sides by $\mathbf{s}_1$, we have

$$\begin{aligned}&\left(\boldsymbol{\varphi}_1(1:3) - \mathbf{s}_1\right)\odot\left(\boldsymbol{\varphi}_1(1:3) - \mathbf{s}_1\right) \\ &\quad\approx \left(\mathbf{u}^o - \mathbf{s}_1\right)\odot\left(\mathbf{u}^o - \mathbf{s}_1\right) + 2\left(\mathbf{u}^o - \mathbf{s}_1\right)\odot\Delta\boldsymbol{\varphi}_1(1:3).\end{aligned} \tag{35}$$

The fourth to the sixth element in $\boldsymbol{\varphi}_1$ denote the velocity estimates of the source. Combing with the position estimates, we have

$$\begin{aligned}&\left(\boldsymbol{\varphi}_1(1:3) - \mathbf{s}_1\right)\odot\left(\boldsymbol{\varphi}_1(4:6) - \dot{\mathbf{s}}_1\right) \\ &\quad\approx \left(\mathbf{u}^o - \mathbf{s}_1\right)\odot\left(\dot{\mathbf{u}}^o - \dot{\mathbf{s}}_1\right) + \left(\dot{\mathbf{u}}^o - \dot{\mathbf{s}}_1\right)\odot\Delta\boldsymbol{\varphi}_1(1:3) \\ &\qquad + \left(\mathbf{u}^o - \mathbf{s}_1\right)\odot\Delta\boldsymbol{\varphi}_1(4:6).\end{aligned} \tag{36}$$

Expressing the seventh element in $\boldsymbol{\varphi}_1$ as $\boldsymbol{\varphi}_1(7) = c^o\hat{r}_1^o + \Delta\boldsymbol{\varphi}_1(7)$, after squaring both sides and using $\boldsymbol{\varphi}_1(8) \approx c^{o2}$ yields

$$\boldsymbol{\varphi}_1(7)^2 \approx \boldsymbol{\varphi}_1(8)\left\|\mathbf{u}^o - \mathbf{s}_1\right\|^2 + 2c^o\hat{r}_1^o\Delta\boldsymbol{\varphi}_1(7). \tag{37}$$

Multiplying $\boldsymbol{\varphi}_1(7)$ and $\boldsymbol{\varphi}_1(9)$ gives

$$\begin{aligned}\boldsymbol{\varphi}_1(7)\boldsymbol{\varphi}_1(9) &= \left(c^o\hat{r}_1^o + \Delta\boldsymbol{\varphi}_1(7)\right)\left(c^o\hat{\dot{r}}_1^o + \Delta\boldsymbol{\varphi}_1(9)\right) \\ &\approx \boldsymbol{\varphi}_1(8)\left(\mathbf{u}^o - \mathbf{s}_1\right)^T\left(\dot{\mathbf{u}}^o - \dot{\mathbf{s}}_1\right) \\ &\quad + c^o\hat{\dot{r}}_1^o\Delta\boldsymbol{\varphi}_1(7) + c^o\hat{r}_1^o\Delta\boldsymbol{\varphi}_1(9).\end{aligned} \tag{38}$$

From (35) to (38), we construct the following matrix equation

$$\boldsymbol{\varepsilon}_2 = \mathbf{h}_2 - \mathbf{G}_2\boldsymbol{\varphi}_2^o, \tag{39}$$

where

$$\boldsymbol{\varphi}_2^o = \begin{bmatrix} (\mathbf{u}^o - \mathbf{s}_1) \odot (\mathbf{u}^o - \mathbf{s}_1) \\ (\mathbf{u}^o - \mathbf{s}_1) \odot (\dot{\mathbf{u}}^o - \dot{\mathbf{s}}_1) \\ c^{o2} \end{bmatrix},$$

$$\mathbf{h}_2 = \begin{bmatrix} (\boldsymbol{\varphi}_1(1:3) - \mathbf{s}_1) \odot (\boldsymbol{\varphi}_1(1:3) - \mathbf{s}_1) \\ (\boldsymbol{\varphi}_1(1:3) - \mathbf{s}_1) \odot (\boldsymbol{\varphi}_1(4:6) - \dot{\mathbf{s}}_1) \\ \boldsymbol{\varphi}_1(7)^2 \\ \boldsymbol{\varphi}_1(7)\boldsymbol{\varphi}_1(9) \\ \boldsymbol{\varphi}_1(8) \end{bmatrix},$$

$$\mathbf{G}_2 = \begin{bmatrix} \mathbf{I}_{3\times3} & \mathbf{0}_{3\times3} & \mathbf{0}_{3\times1} \\ \mathbf{0}_{3\times3} & \mathbf{I}_{3\times3} & \mathbf{0}_{3\times1} \\ \boldsymbol{\varphi}_1(8)\mathbf{1}_{1\times3} & \mathbf{0}_{1\times3} & 0 \\ \mathbf{0}_{1\times3} & \boldsymbol{\varphi}_1(8)\mathbf{1}_{1\times3} & 0 \\ \mathbf{0}_{1\times3} & \mathbf{0}_{1\times3} & 1 \end{bmatrix}.$$

On the left side of (39), the noise vector $\boldsymbol{\varepsilon}_2$ is defined as

$$\boldsymbol{\varepsilon}_2 = \mathbf{B}_2\Delta\boldsymbol{\varphi}_1, \tag{40}$$

where

$$\mathbf{B}_2 = \begin{bmatrix} 2diag([\mathbf{u}^o - \mathbf{s}_1]) & \mathbf{0}_{3\times3} & \mathbf{0}_{3\times3} \\ diag([\dot{\mathbf{u}}^o - \dot{\mathbf{s}}_1]) & diag([\mathbf{u}^o - \mathbf{s}_1]) & \mathbf{0}_{3\times3} \\ \mathbf{0}_{3\times3} & \mathbf{0}_{3\times3} & \mathbf{D}_2 \end{bmatrix},$$

$$\mathbf{D}_2 = \begin{bmatrix} 2c^o\hat{r}_1^o & 0 & 0 \\ c^o\hat{\dot{r}}_1^o & 0 & c^o\hat{r}_1^o \\ 0 & 1 & 0 \end{bmatrix}.$$

We would like to remark that the noise vector in (40) is only related with the estimation error resulted from the first stage. This is because of the premeditated applying of Taylor approximation to $r_1^o$ and $\dot{r}_1^o$, which is expressed in (19) and (23), respectively. As a result, $\boldsymbol{\varepsilon}_2$ is independent of the SN parameter uncertainties $\Delta\boldsymbol{\beta}$. It is different from the second-stage error vector derived in [13], in which the error vector depends on both the first stage estimation error and the SN parameter uncertainties. It has also been observed in [27] that the localization performance is improved by expanding $r_1^o$ and $\dot{r}_1^o$ in the first stage. Furthermore, the simplified structure of $\boldsymbol{\varepsilon}_2$ gives a concise expression of its covariance matrix as

$$cov(\boldsymbol{\varepsilon}_2) = E\left(\boldsymbol{\varepsilon}_2\boldsymbol{\varepsilon}_2^T\right) = \mathbf{B}_2\left(\mathbf{G}_1^T\mathbf{W}_1\mathbf{G}_1\right)^{-1}\mathbf{B}_2^T, \tag{41}$$

where (34) is used.

The WLS solution to (39) is given by

$$\boldsymbol{\varphi}_2 = \left(\mathbf{G}_2^T\mathbf{W}_2\mathbf{G}_2\right)^{-1}\mathbf{G}_2^T\mathbf{W}_2\mathbf{h}_2, \tag{42}$$

where $\mathbf{W}_2$ is the weighting matrix defined as

$$\mathbf{W}_2 = (cov(\boldsymbol{\varepsilon}_2))^{-1} = \mathbf{B}_2^{-T}\left(\mathbf{G}_1^T\mathbf{W}_1\mathbf{G}_1\right)\mathbf{B}_2^{-1}. \tag{43}$$

Subtracting both sides of (42) by the true value $\boldsymbol{\varphi}_2^o$ gives the estimation error of the second stage

$$\Delta\boldsymbol{\varphi}_2 = \boldsymbol{\varphi}_2 - \boldsymbol{\varphi}_2^o = \left(\mathbf{G}_2^T\mathbf{W}_2\mathbf{G}_2\right)^{-1}\mathbf{G}_2^T\mathbf{W}_2\left(\mathbf{h}_2 - \mathbf{G}_2\boldsymbol{\varphi}_2^o\right) = \left(\mathbf{G}_2^T\mathbf{W}_2\mathbf{G}_2\right)^{-1}\mathbf{G}_2^T\mathbf{W}_2\boldsymbol{\varepsilon}_2. \tag{44}$$

Then, the covariance matrix of the solution $\boldsymbol{\varphi}_2$ can be approximated by

$$cov(\boldsymbol{\varphi}_2) = \left(\mathbf{G}_2^T\mathbf{W}_2\mathbf{G}_2\right)^{-1}. \tag{45}$$

Finally, the source position and velocity, and the sound propagation speed estimates are given by

$$\begin{aligned} \mathbf{u} &= \Pi\sqrt{\boldsymbol{\varphi}_2(1:3)} + \mathbf{s}_1, \\ \dot{\mathbf{u}} &= \boldsymbol{\varphi}_2(4:6)./(\mathbf{u} - \mathbf{s}_1) + \dot{\mathbf{s}}_1, \\ c &= \sqrt{\boldsymbol{\varphi}_2(7)}, \end{aligned} \tag{46}$$

where $\mathbf{\Pi} = diag\left(\left[sign\left(\boldsymbol{\varphi}_1(1:3) - \mathbf{s}_1\right)\right]\right)$ is used to avoid the sign ambiguity caused by the square root operation.

Note that the weighting matrix $\mathbf{W}_2$ is dependent on $\mathbf{u}^o$, $\dot{\mathbf{u}}^o$ and $c^o$ through $\mathbf{B}_2$. Similar to the calculation of $\mathbf{W}_1$, $\mathbf{W}_2$ can also be updated in an iterative fashion. The true values in $\mathbf{W}_2$ are first approximated by the values in $\boldsymbol{\varphi}_1$ and then updated by the values in (46). Also, we find that iterating one or two times leads to an good solution that meets the CRLB performance.

We summarize the prototype of our proposed estimator in Algorithm 1.

**Algorithm 1** The Proposed Estimator

**Input:** SN parameters, TDOA and FDOA measurements and error covariance matrices.

*First stage processing:*
1: *Initialization:* $\mathbf{W}_1 = \mathbf{Q}_{\boldsymbol{\alpha}}^{-1}$.
2: **For** $l = 1$ **to** $N_{iter}$ ($N_{iter}$ is the number of iterations)
3: computing $\boldsymbol{\varphi}_1$ from (30);
4: substituting the estimates from $\boldsymbol{\varphi}_1$ in (32) to update $\mathbf{W}_1$;

5: **end For**
*Second stage processing:*
6: Computing $cov\left(\boldsymbol{\varphi}_1\right)$ using (34).
7: Using $\boldsymbol{\varphi}_1$ to calculate $\mathbf{B}_2$ and obtaining $\mathbf{W}_2$ using (43).
8: **For** $l = 1$ **to** $N_{iter}$
9: computing $\boldsymbol{\varphi}_2$ from (42);
10: applying (46) to generate the estimates;
11: substituting the estimates from (46) in $\mathbf{B}_2$ and updating $\mathbf{W}_2$ using (43) accordingly;
12: **end For**

**Output:** the source position and velocity, and the sound propagation speed estimates.

## V. PERFORMANCE ANALYSIS

In this section, the performance of the proposed two stage algorithm shall be evaluated. We compare the theoretical covariance matrix of the solution given by (46) with the CRLB derived in Section (III). Note that the following analysis is valid under the small noise condition. Hence, it shows the asymptotically performance of the proposed algorithm.

We define the parameter vector $\boldsymbol{\xi}^o = \left[ \left( \boldsymbol{\theta}^o \right)^T, c^o \right]^T$ and its estimate $\boldsymbol{\xi}$ is given by (46). According to the definition of $\boldsymbol{\varphi}_2^o$ below (39) and taking the differential gives

$$\Delta \boldsymbol{\xi} = \boldsymbol{\xi} - \boldsymbol{\xi}^o = \left[ \Delta \mathbf{u}^T \quad \Delta \dot{\mathbf{u}}^T \quad \Delta c \right]^T = \mathbf{B}_3^{-1} \Delta \boldsymbol{\varphi}_2, \quad (47)$$

where

$$\mathbf{B}_3 = \begin{bmatrix} 2diag\left( \mathbf{u}^o - \mathbf{s}_1 \right) & \mathbf{0}_{3\times 3} & \mathbf{0}_{3\times 1} \\ diag\left( \dot{\mathbf{u}}^o - \dot{\mathbf{s}}_1 \right) & diag\left( \mathbf{u}^o - \mathbf{s}_1 \right) & \mathbf{0}_{3\times 1} \\ \mathbf{0}_{1\times 3} & \mathbf{0}_{1\times 3} & 2c^o \end{bmatrix}. \quad (48)$$

The bias of the solution estimate is given by taking expectation of (47). Under the small noise assumption, $\Delta\boldsymbol{\xi}$ is linearly related to $\left[ \Delta\boldsymbol{\alpha}^T, \Delta\boldsymbol{\beta}^T \right]^T$ through (31), (33), (40), (44) and (47). Since $\left[ \Delta\boldsymbol{\alpha}^T, \Delta\boldsymbol{\beta}^T \right]^T$ is zero mean, $\Delta\boldsymbol{\xi}$ is also zero mean, which implies that the estimate is unbiased. The covariance matrix of $\boldsymbol{\xi}$ can be expressed as

$$cov\left( \boldsymbol{\xi} \right) = E\left( \Delta\boldsymbol{\xi} \Delta\boldsymbol{\xi}^T \right) = \mathbf{B}_3^{-1} cov\left( \boldsymbol{\varphi}_2 \right) \mathbf{B}_3^{-T}. \quad (49)$$

After substituting (45), (43) and (32) and applying the matrix inversion Lemma [36], (49) can be written as

$$\begin{aligned} cov(\boldsymbol{\xi})^{-1} &= \mathbf{G}_3^T \mathbf{Q}_{\boldsymbol{\alpha}}^{-1} \mathbf{G}_3 \\ &\quad - \mathbf{G}_3^T \mathbf{Q}_{\boldsymbol{\alpha}}^{-1} \mathbf{G}_4 \left( \mathbf{Q}_{\boldsymbol{\beta}}^{-1} + \mathbf{G}_4^T \mathbf{Q}_{\boldsymbol{\alpha}}^{-1} \mathbf{G}_4 \right)^{-1} \mathbf{G}_4^T \mathbf{Q}_{\boldsymbol{\alpha}}^{-1} \mathbf{G}_3, \end{aligned} \quad (50)$$

where $\mathbf{G}_3 = \mathbf{B}_1^{-1}\mathbf{G}_1\mathbf{B}_2^{-1}\mathbf{G}_2\mathbf{B}_3$ and $\mathbf{G}_4 = \mathbf{B}_1^{-1}\mathbf{D}_1$. According to (7), the CRLB inverse of $\boldsymbol{\xi}^o$ is given by

$$CRLB(\boldsymbol{\xi}^o)^{-1} = \begin{bmatrix} \mathbf{X}_{11} & \mathbf{X}_{12} \\ \mathbf{X}_{12}^T & \mathbf{X}_{22} \end{bmatrix} - \begin{bmatrix} \mathbf{X}_{13} \\ \mathbf{X}_{23} \end{bmatrix} \mathbf{X}_{33}^{-1} \left[ \mathbf{X}_{13}^T \quad \mathbf{X}_{23}^T \right]. \quad (51)$$

Comparing (50) with (51) shows that they are of the same functional form. After substituting (9) into (7), to arrive at $cov(\boldsymbol{\xi})^{-1} \approx CRLB(\boldsymbol{\xi}^o)^{-1}$, we find that we just need to prove that

1) $\mathbf{G}_3 \approx \pm \left[ \partial\boldsymbol{\alpha}^o / \partial\boldsymbol{\theta}^o \quad \partial\boldsymbol{\alpha}^o / \partial c^o \right]$,
2) $\mathbf{G}_4 \approx \pm \partial\boldsymbol{\alpha}^o / \partial\boldsymbol{\beta}^o$.

To facilitate the proof derivation, we shall establish the following three additional conditions besides the small noise condition (i)

1) $r_1^o \ll r_i^o, i = 2, \ldots, M$,
2) $\dot{r}_i^o / r_i^o \approx 0, i = 1, \ldots, M$,
3) $t_{i1} / r_i^o \approx 0, i = 1, \ldots, M$.

The first condition indicates that the SN nearest the source should be chosen as the reference node. The second condition states that the moving velocity relative to the range is close to 0. The third condition requires that the signal propagation time relative to range approximates 0. In the underwater environment, objects usually move slowly (several meters per second) and the sound propagation speed is about 1500 m/s. Therefore, conditions (ii) and (iii) are valid for a localization area with radius upto several kilometers. Note that the conditions (i) and (iii) are different from the conditions in [13], thus it is necessary to rederive the proof process. Appendix C shows that the equations $\mathbf{G}_3 \approx \left[ \partial\boldsymbol{\alpha}^o / \partial\boldsymbol{\theta}^o \; \partial\boldsymbol{\alpha}^o / \partial c^o \right]$ and $\mathbf{G}_4 \approx \partial\boldsymbol{\alpha}^o / \partial\boldsymbol{\beta}^o$ are approximately hold under the conditions established. Consequently, we conclude that the estimation accuracy provided by (46) attains the CRLB under small Gaussian noise and the conditions (i-iii).

To end this section, we would like to make some remarks on the condition (i). In order to satisfy this condition for a fixed $r_i^o$, $i \neq 1$, it is better to make $r_1^o$ as small as possible. However, when $r_1^o$ is close to 0, the vector $(\mathbf{u}^o - \mathbf{s}_1)$ is close to 0 as well. This destroys the nonsingularity of the matrix $\mathbf{B}_2$, which renders the calculation of the weighting matrix of the second stage to be intractable. Hence, in practice, we prefer to choose a moderate value of $r_j^o$ from the ranges $r_i^o$, $i = 1, \ldots, M$ that satisfying $0 < r_j^o < r_i^o, i \neq j$ to be the reference range $r_1^o$.

## VI. SIMULATION RESULTS

In this section, the performance of the proposed method is evaluated through Monte Carlo simulations and compared with the CRLB and three analogous existing methods, i.e., the three-step solution [30], the WLS solution [27] and the SDP solution [25]. Note that none of the localization models in these works is identical to ours. We choose these three existing methods due to their comparability and novelty. In each simulation trail, without otherwise specified, the true sound propagation speed $c^o$ is randomly drawn from the range [1400, 1600] m/s. The other simulation settings are same as the one specified in Section III-B. The mean square error (MSE) is used to evaluate the estimation performance of parameter $\boldsymbol{\xi}$. The MSE is computed as $MSE(\boldsymbol{\xi}) = \sum_{i=1}^{N} \left\| \boldsymbol{\xi}^i - \boldsymbol{\xi}^o \right\|^2 / N$, where $\boldsymbol{\xi}^i$ is the estimate of $\boldsymbol{\xi}^o$ at ensemble $i$. Note that we present all the MSE results in decibels (dB) with reference level 1 $m^2$ or 1 $m^2/s^2$. The results are shown with the values taken logarithm of base 10. The reason for using dB instead of $m^2$ or $m^2/s^2$ as the unit is that the former can show the results in a large scale and make the curve more smooth. The number of ensemble runs $N$ is 1000.

### A. IMPACT OF THE MEASUREMENT NOISE

Fig. 3 illustrates the performance of the proposed method as the measurement noise variance increases. The SN uncertainty $\sigma_s$ is fixed to be 1 m. Note that the three-step solution can not perform velocity estimation while the WLS solution and SDP solution can not perform sound propagation speed estimation. In order to emphasize the impact of sound propagation speed error on the localization accuracy, we use $c^o$+20 m/s as the propagation speed estimate of WLS solution and SDP solution. From Fig. 3, the following observations can be made:

1. The proposed method achieves the CRLB accuracy and performs better than other methods for source position,

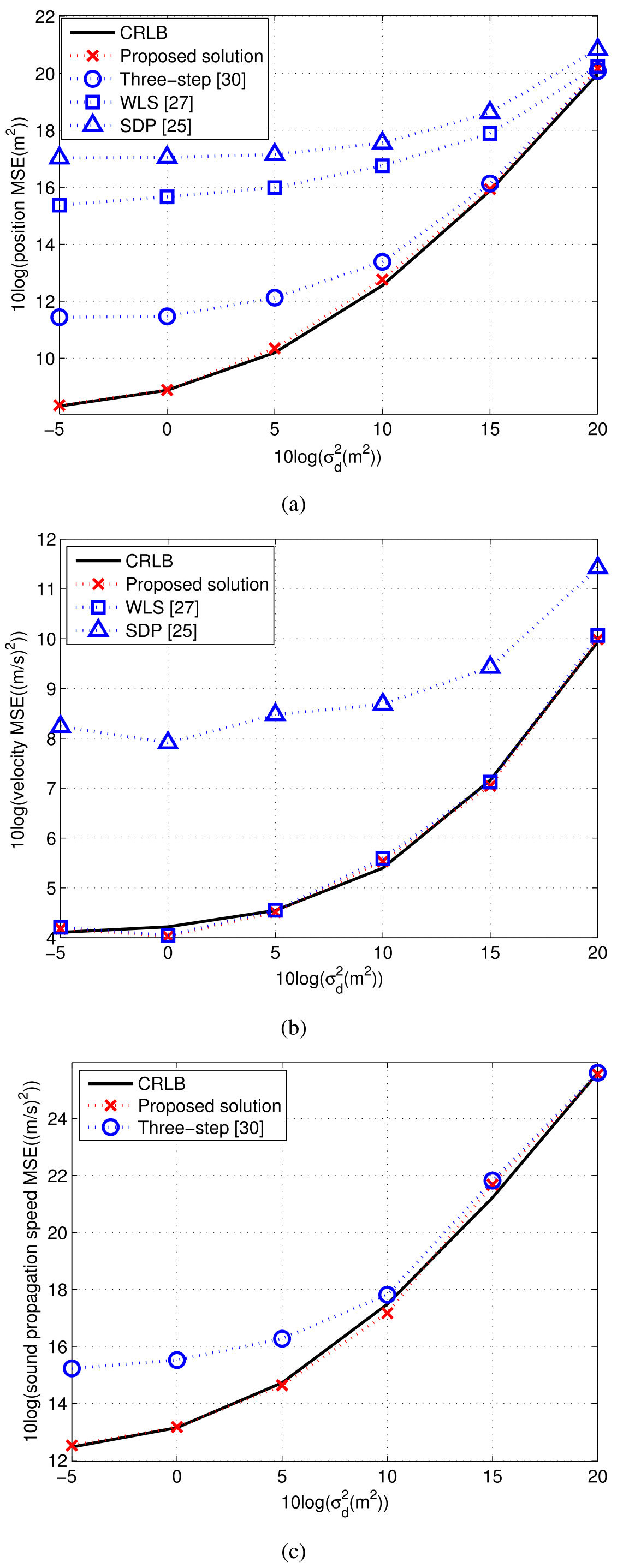

**FIGURE 3.** Impact of the measurement noise on MSE performance of the considered algorithms: (a) source position; (b) source velocity; (c) sound propagation speed.

velocity and sound propagation speed estimates. The superiority of the proposed method over other methods is more significantly for smaller noise conditions.

2. The other methods perform comparable to the proposed method at high noise level. This is mainly because of that the performance bound derived here includes three kinds of unknown parameters, i.e. source position, velocity and sound propagation speed. However, the other methods only consider two kinds of unknown parameters, source position and velocity or source position and sound propagation speed. Thus, the CRLB derived here is not suitable for the other methods. In other words, as the noise variance increases, the CRLB accounting for more unknowns increases more faster. Nevertheless, our proposed method is still able to reach the CRLB and this corroborates the theoretical analysis in Section V.
3. The gap between the three-step method and the proposed method is mainly because the three-step method incorporates no FDOA measurements and takes no SN parameter uncertainties into account. The need to estimate the sound propagation speed is demonstrated by the significant performance gain of the proposed method compared to that of the WLS and SDP solution.
4. Interestingly, we note that the WLS solution provides an excellent velocity estimation which is merely slightly inferior to the proposed method. However, this phenomenon is not observed in the SDP solution. We collect the points of the WLS and proposed solution from Fig. 3(b) into Table 2 to illustrate the difference clearly. This is because that the SDP method uses the SDR procedure which drops the rank-1 constraint. As a result, the SDP method holds a worse performance particularly for small measurement noise condition.

**TABLE 2.** MSEs of velocity for the WLS and proposed solution versus $\sigma_d^2$.

| $\sigma_d^2$ (dB) | −5 | 0 | 5 | 10 | 15 | 20 |
|---|---|---|---|---|---|---|
| WLS (dB) | 4.211 | 4.05 | 4.552 | 5.592 | 7.125 | 10.07 |
| Proposed (dB) | 4.182 | 4.03 | 4.515 | 5.532 | 7.046 | 9.986 |

### B. IMPACT OF THE SN PARAMETER ERROR

Fig. 4 shows the results at different levels of SN parameter error $\sigma_s$ when $\sigma_d$ is set to be 1 m and other simulation parameters remain unchanged. The following observations can be made:

1. As expected, the performance of the proposed method attains the CRLB for all to be estimated parameters. While the three-step approach also follows the trend of CRLB, it starts to sperate from the CRLB for higher SN parameter error. This is a consequence of ignoring the SN parameter uncertainty in deriving the three-step solution.
2. In Fig. 4(a), comparing the proposed method and the one that does not account for the sound propagation speed error when $\sigma_s^2 \leq -5$ dB, the new method achieves an reduction in the source position MSE more than 10 dB.
3. Again, it is observed from Fig. 4(b) that the WLS solution is very close to the CRLB and significantly outperforms the SDP solution for the velocity estimation.

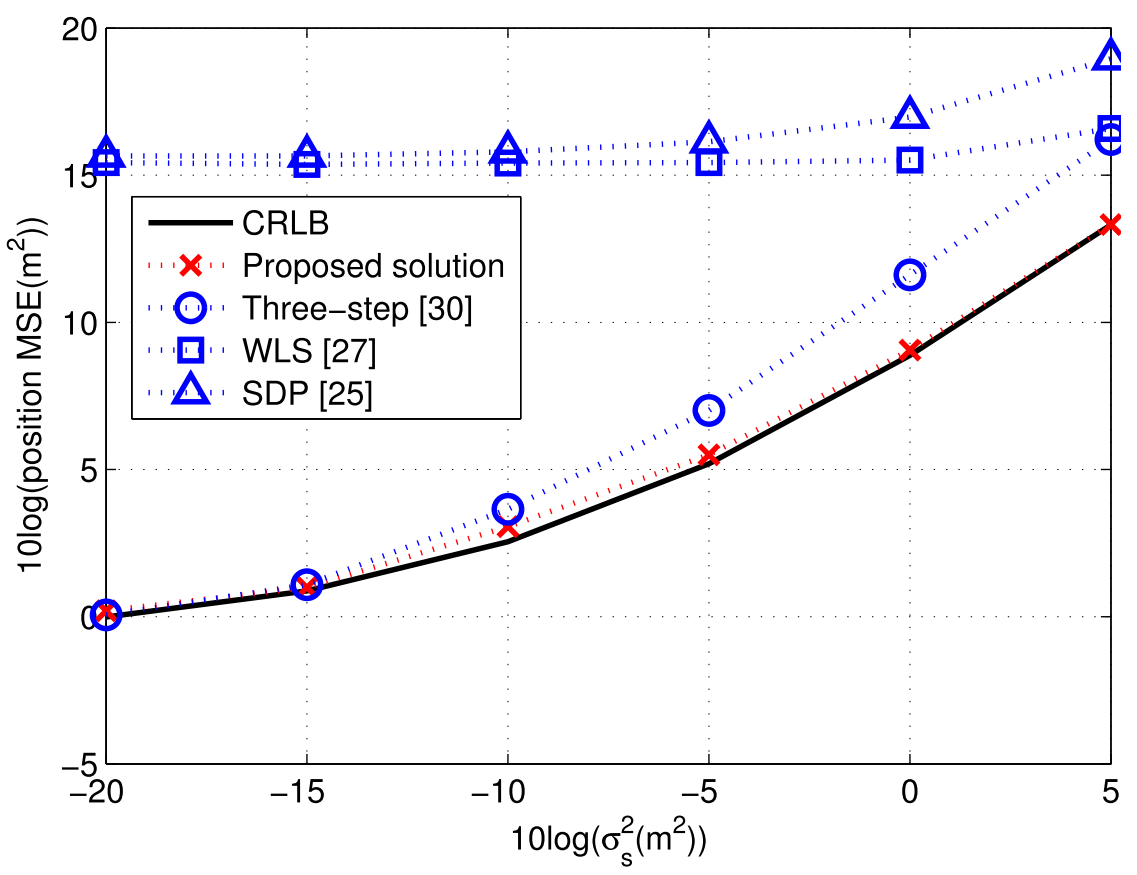

(a)

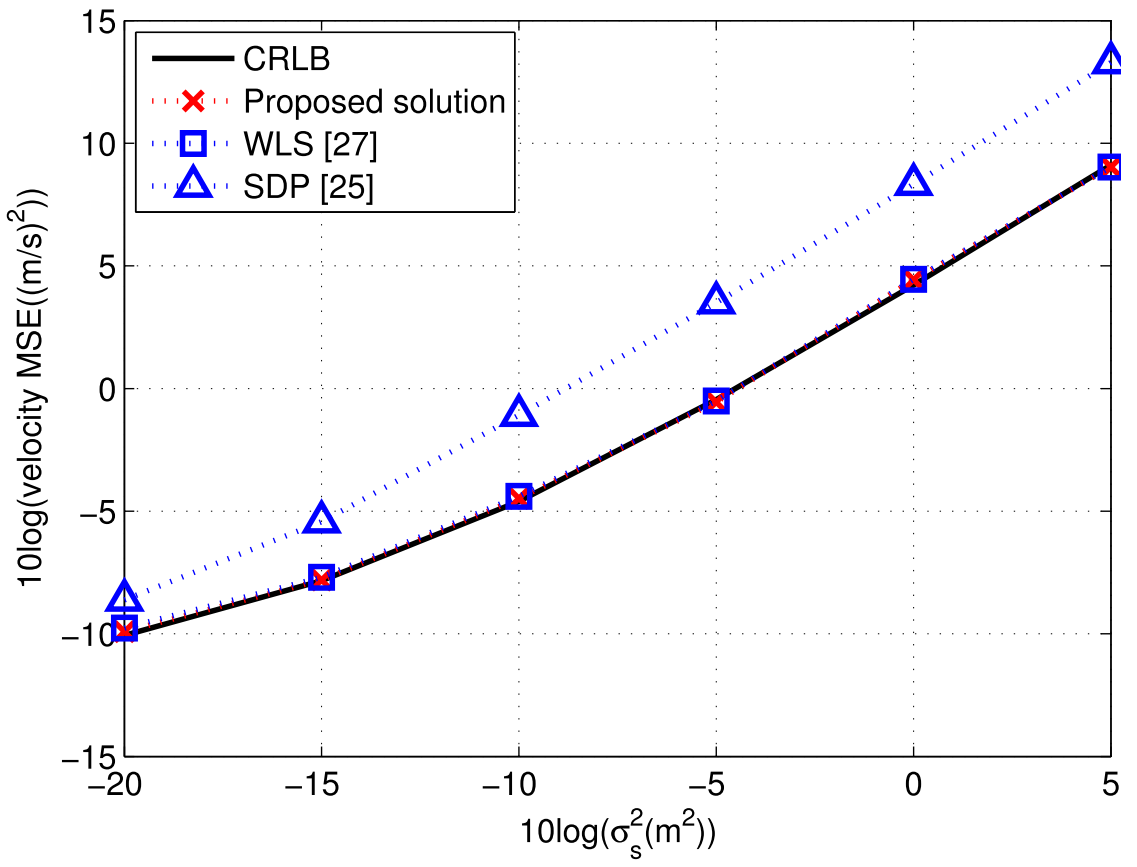

(b)

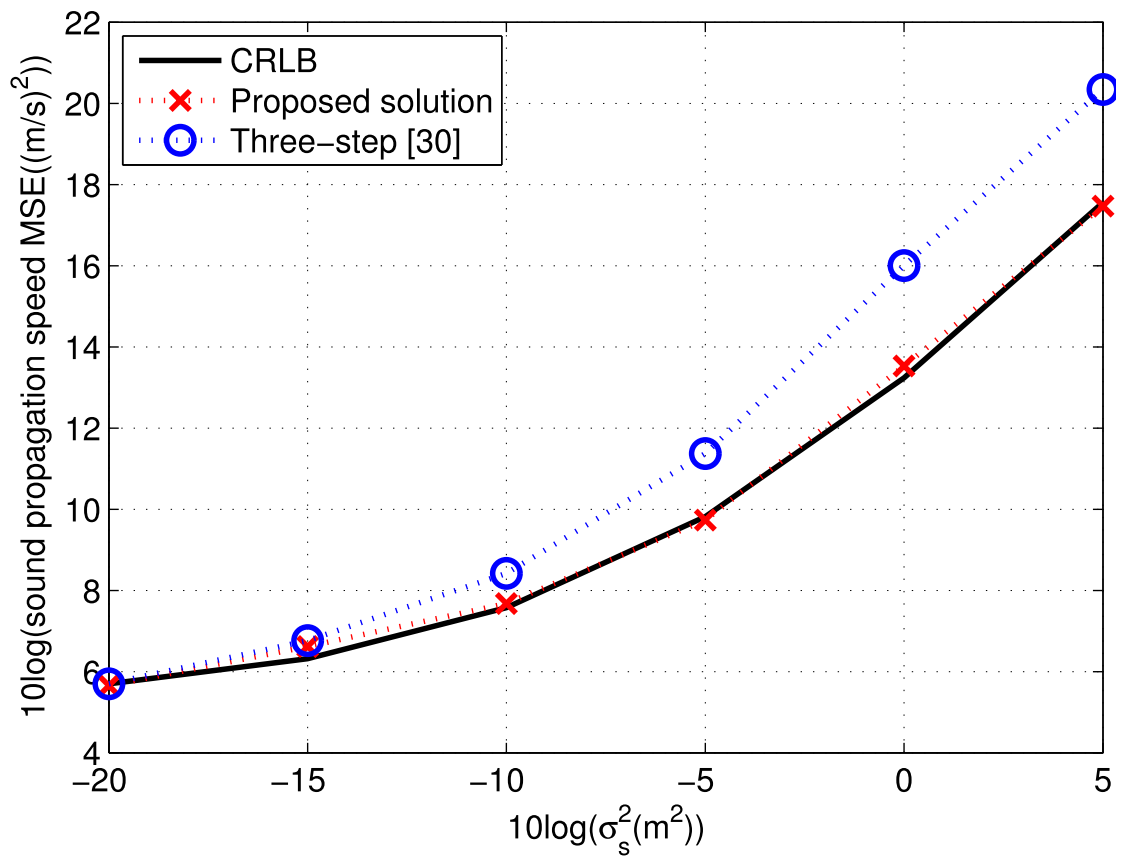

(c)

**FIGURE 4.** Impact of the SN parameter errors on MSE performance of the considered algorithms: (a) source position; (b) source velocity; (c) sound propagation speed.

We also collect the points of the WLS and proposed solution from Fig. 4(b) into Table 3 to illustrate the difference clearly. The reason behind this is the same as in Fig. 3.

4. Moreover, we would like to point out that the values of velocity used in simulation are much smaller than the position value. This scale difference explains why the WLS velocity MSE performance shows to be "much better" than that of its position comparing Fig. 4(a) and Fig. 4(b).

**TABLE 3.** MSEs of velocity for the WLS and proposed solution versus $\sigma_s^2$.

| $\sigma_d^2$ (dB) | −5 | 0 | 5 | 10 | 15 | 20 |
|---|---|---|---|---|---|---|
| WLS (dB) | -9.72 | -7.7 | -4.40 | -0.51 | 4.45 | 9.03 |
| Proposed (dB) | -9.91 | -7.8 | -4.475 | -0.55 | 4.436 | 9.03 |

### C. IMPACT OF INACCURATE SOUND PROPAGATION SPEED

In this simulation, we aim to study the impact of inaccurate sound propagation speed on the considered algorithms. The SN parameter error and the measurement noise variance are set to be 0 dB and −5 dB respectively. The nominal sound speed is set as $c^o = 1490$ m/s and the actual sound speed is $c = c^o + \Delta c$ m/s, where $\Delta c \in [-70, 70]$ m/s. Note that the range of $\Delta c$ adopted here is based on a widely used model that bounds propagation speed underwater between 1420 m/s and 1560 m/s [16]. As it can be seen from Fig. 5, with an increase of the sound speed error (both in positive or negative direction), the performances of the algorithms which have no regard for the sound speed variation get worse. However, it has no effect on the three-step method and our proposed method. This illustrates the importance of taking the inaccurate sound speed into account while designing the underwater localization algorithms. We note that MSE performance of the WLS method is better than that of our proposed method and even the CRLB when there is no sound speed error, i.e. $\Delta c = 0$ m/s. This is mainly due to the fact that the main strength of our proposed method is to deal with the localization situation where sound propagation speed is unknown, and the CRLB derived here incorporates three kinds of unknown parameters, i.e. source position, velocity and sound propagation speed. Therefore, this CRLB is not suitable for the WLS method which assumes that the signal propagation speed is perfectly known. Nevertheless, the root MSE performance gap between the WLS method and our proposed one is less than 1 m (0.29 m in Fig. 5) at the point $\Delta c = 0$ m/s, which can be safely ignored.

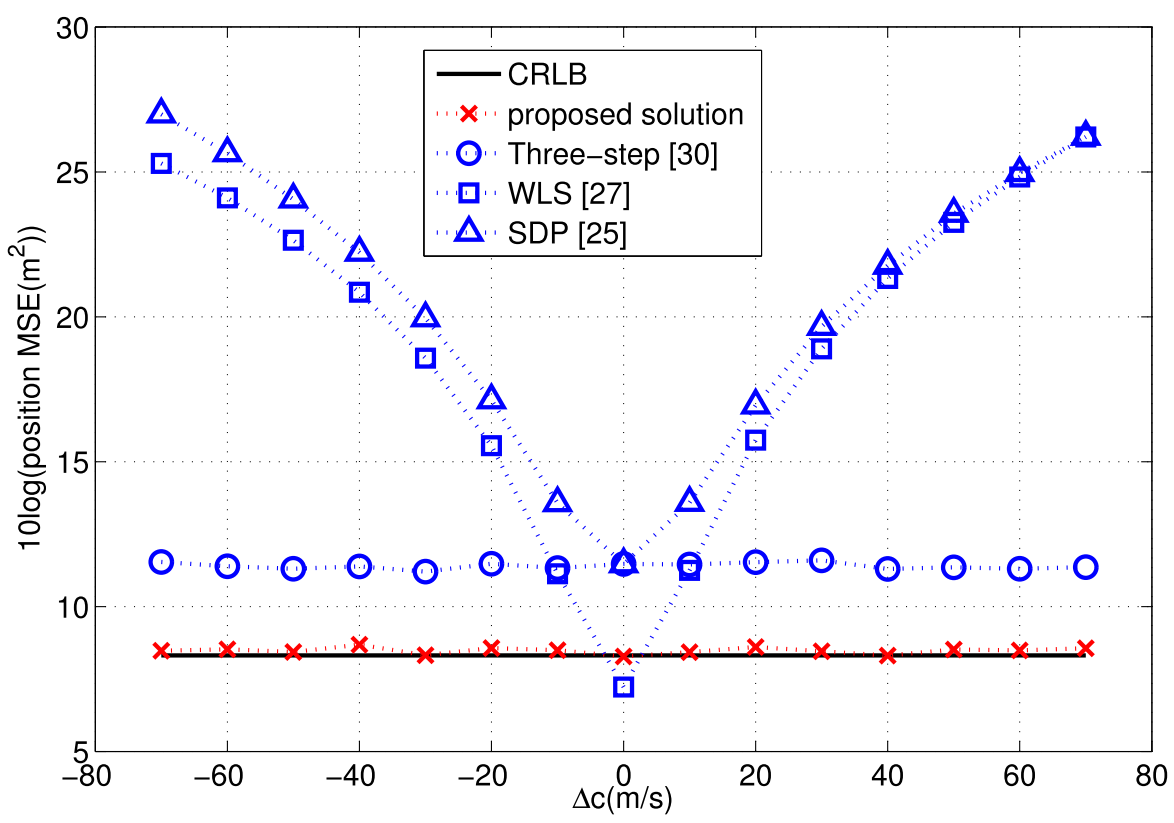

**FIGURE 5.** Impact of the inaccurate sound propagation speed on MSE performance of the considered algorithms, where $\Delta c$ is the error to the nominal sound speed.

### D. COMPARISON OF THE TREATMENTS OF THE WEIGHTING MATRIX $W_1$

In real life, the error covariance matrices $\mathbf{Q}_\alpha$ and $\mathbf{Q}_\beta$ can be obtained by pre-calibration. However, they might also be unavailable due to the pre-calibration cost or other reasons. For each case, we suggest corresponding treatment of the weighting matrix $\mathbf{W}_1$ in Section IV-A. In this subsection, we conduct simulations to study the impacts of these treatments on the localization performance of our proposed solution. From Fig. 6, the following observations can be made:

1. As predicted, the performance of the proposed solution degrades without the error covariance information especially at the small noise region. To explain this, the second stage of the proposed method is performed based on its first stage solution. Under a small noise, the first stage can be very accurate with the aid from the error covariance information, thus the approximations used in the second stage are comparatively reasonable. Therefore, more accurate estimation is achieved. However, if the error covariance information is unavailable in the first stage, the weighting matrix is not optimal in the WLS sense, which leads to a worse first stage solution and consequently introduces lager approximation errors in the second stage processing. Nevertheless, we would like to point out that the performance degradation of our proposed method without the error covariance information is acceptable. As observed from Fig. 6, the largest performance gap happened at $\sigma_d^2 = -5$ dB, the root MSEs of the proposed method with and without error covariance information is 2.66 m and 4.1 m, respectively. The difference in root RMSE is 1.44 m which can be tolerated by most underwater applications.

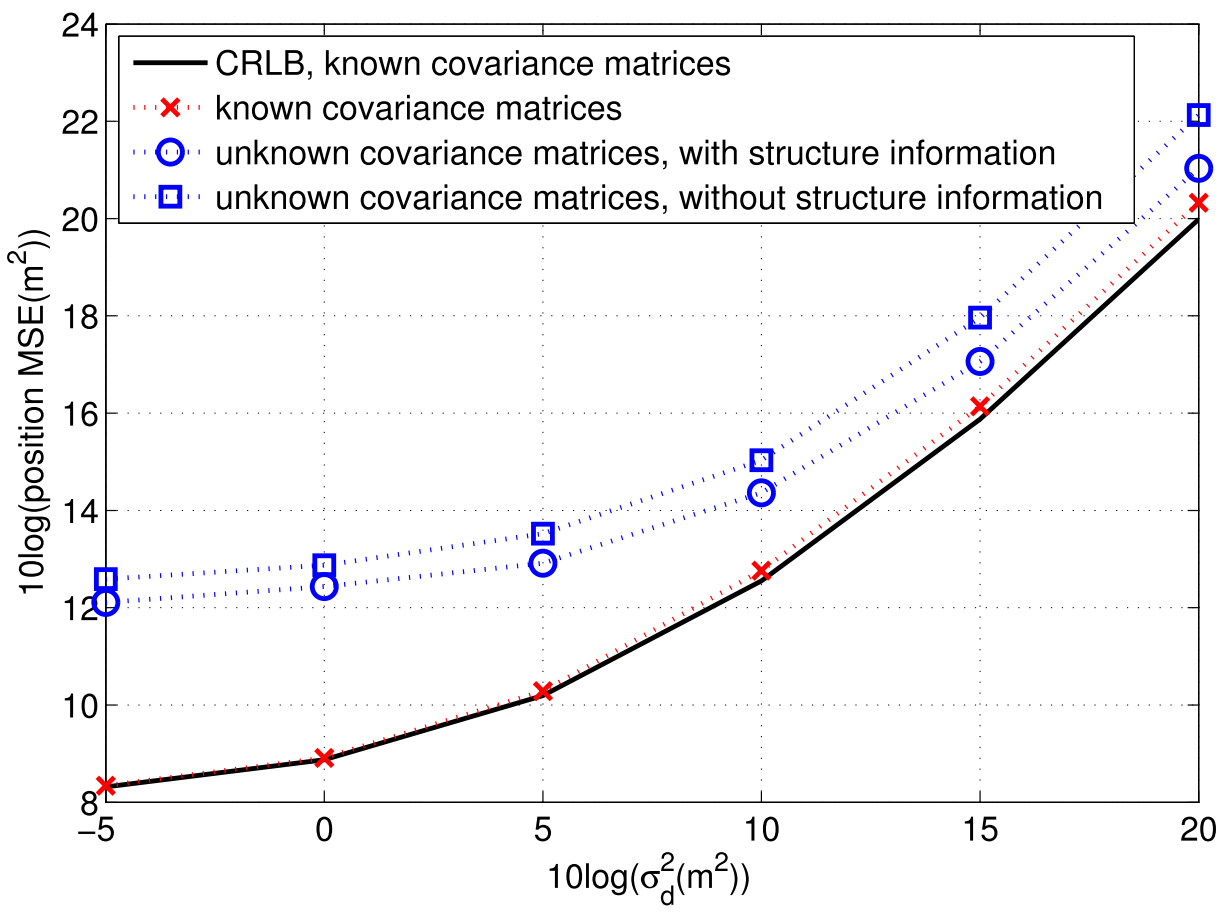

**FIGURE 6. Comparison of the treatments of the weighting matrix $\mathbf{W}_1$ for the proposed solution.**

2. Even without the knowledge of the error covariance information, exploiting the weighting matrix structure information (i.e. the matrices $\mathbf{B}_1$ and $\mathbf{D}_1$ in (32).) provides an improved estimation performance over the one that ignoring the structure information. The benefit of using the structure information increases as the measurement noise increases. To explain this, we need to recall that the ignoring of the second order noise term and the Taylor approximation used in our proposed solution might become inaccurate under a large measurement noise, thus making our method not as effective as it is under a small measurement noise. Under this circumstance, an inaccurate weighting matrix can also exacerbate the impact of a large measurement noise, which makes the performance of our proposed method worse. The using of structure information undoubtedly compensates the inaccurate weighting matrix and provides the benefit observed in Fig. 6.

To conclude the impact of unknown error covariance matrices on our proposed method, we claim that the performance degrades without the error covariance information but within an acceptable level, i.e. several meters. What's more, the structure information of the weighting matrix should be considered to alleviate this degradation.

### E. COMPLEXITY ANALYSIS

Regarding the computational complexity of the different methods, we provide both analysis and simulation results. We assume that the complexity of a matrix multiplication operation (one $n \times m$ matrix and one $m \times p$ matrix) is $O(nmp)$, and that of a matrix inversion operation (one $n \times n$ matrix) is $O(n^3)$. The computation overhead of the proposed solution includes two parts, i.e. the first stage calculation and the second stage calculation. However, it is noteworthy that the scale of the second stage calculation is much less than that of the first stage. As a result, we only focus on the computation complexity of the first stage. It is easy to drive that the complexity of computing (30) is $O(M^2)$. The updating of the weighting matrix $\mathbf{W}_1$ in (32) requires a computation load of $O(M^3)$. Thus, the complexity of the proposed solution can be approximated as $O(M^3)$. As for the three-step solution, its complexity is $O(M^3)$. Comparing with the proposed method, a reduction of computation cost is expected due to the absence of FDOA measurements. Similarly, the complexity of the WLS solution can be expressed as $O(M^3)$. The worst-case complexity of solving the SDP solution is $O((u^3 + u^2v^2 + uv^3)v^{0.5})$ [38], where $u$ is the number of equality constrains, and $v$ is the problem size. In the compared SDP solution, $u = 4$ and $v = 9$. The complexity of computing the weighting matrix in [25] is roughly $O(M)$. The number of iterations is bounded by $O(\sqrt{M} ln(1/\xi))$, where $\xi$ is the iteration tolerance. Therefore, the total complexity of the SDP solution is $O(\sqrt{M} ln(1/\xi)(u^3 + u^2v^2 + uv^3)v^{0.5} + M)$.

**TABLE 4.** Average CPU computational time per estimation for different algorithms.

| Algorithm | Average CPU Time (ms) |
|---|---|
| Proposed solution | 1.7 |
| Three-step [30] | 0.6 |
| WLS [27] | 1.9 |
| SDP [25] | 370 |

We give a comparison of the average CPU computational time per estimation in Table 4. The algorithms are implemented in MATLAB R2013a on a lenovo T440 (Processor 4 GHz Intel Core i5, Memory 4GB). We can see that the Three-step method requires the least CPU time among all the considered algorithms while the SDP method requires the most CPU time due to its iteration manner. The required computation overheads of the WLS method and the proposed method are comparable. Comparing with the three-step method, our proposed algorithm is able to provide the velocity estimate and more accurate position estimate with a lightly increased computational load.

## VII. CONCLUSION AND FUTURE WORK

In this paper, we have proposed an algebraic solution for estimating a moving source's position and velocity using TDOA and FDOA measurements in underwater environment. Due to the harsh underwater environment, we assume the measured SN parameters are not accurate and subject to a Gaussian distributed error. Moreover, the sound propagation speed is also assumed to be unknown, and we estimate it together with the unknown source parameters. The proposed solution introduces nuisance variables to obtain a set of pseudo-linear equations in the first stage and improves the estimation accuracy in the second stage by exploiting the nonlinear relationship among nuisance variables. We have derived the CRLB for the localization problem and it shows that the unknown sound propagation speed affects the measurement noise covariance matrix by an orthogonal projection of its square root inverse. The performance of the proposed method is shown in theory and by simulations to reach the CRLB accuracy for the source position, velocity and sound propagation speed estimates at moderate noise level before the thresholding effect occurs.

The study in this correspondence only illustrates the performance of the proposed method by simulations due to the experimental equipment limitations. In the future, we plan to implement it in underwater testbeds and explore its application in real oceanic environment. Another possible extension of the current work is to consider the link-wise sound propagation speed in the localization scenario.

## APPENDIX A
## PARTIAL DERIVATIVES IN (9)

By definition, we have $\boldsymbol{\alpha}^o = \left[\frac{1}{c^o}(\mathbf{r}^o)^T, \frac{1}{c^o}(\dot{\mathbf{r}}^o)^T\right]^T$ and its derivative with respect to (w.r.t.) the parameter vector $\boldsymbol{\phi}^o$ can be calculated by

$$\frac{\partial \boldsymbol{\alpha}^o}{\partial \boldsymbol{\phi}^o} = \left[\frac{\partial \boldsymbol{\alpha}^o}{\partial \boldsymbol{\theta}^o} \quad \frac{\partial \boldsymbol{\alpha}^o}{\partial c^o} \quad \frac{\partial \boldsymbol{\alpha}^o}{\partial \boldsymbol{\beta}^o}\right] = \begin{bmatrix} \frac{\partial(\mathbf{r}^o/c^o)}{\partial \boldsymbol{\theta}^o} & \frac{\partial(\mathbf{r}^o/c^o)}{\partial c^o} & \frac{\partial(\mathbf{r}^o/c^o)}{\partial \boldsymbol{\beta}^o} \\ \frac{\partial(\dot{\mathbf{r}}^o/c^o)}{\partial \boldsymbol{\theta}^o} & \frac{\partial(\dot{\mathbf{r}}^o/c^o)}{\partial c^o} & \frac{\partial(\dot{\mathbf{r}}^o/c^o)}{\partial \boldsymbol{\beta}^o} \end{bmatrix}, \quad (52)$$

where

$$\frac{\partial(\mathbf{r}^o/c^o)}{\partial \boldsymbol{\theta}^o} = \left[\frac{\partial(\mathbf{r}^o/c^o)}{\partial \mathbf{u}^o} \quad \frac{\partial(\mathbf{r}^o/c^o)}{\partial \dot{\mathbf{u}}^o}\right],$$
$$\frac{\partial(\dot{\mathbf{r}}^o/c^o)}{\partial \boldsymbol{\theta}^o} = \left[\frac{\partial(\dot{\mathbf{r}}^o/c^o)}{\partial \mathbf{u}^o} \quad \frac{\partial(\dot{\mathbf{r}}^o/c^o)}{\partial \dot{\mathbf{u}}^o}\right],$$
$$\frac{\partial(\mathbf{r}^o/c^o)}{\partial \boldsymbol{\beta}^o} = \left[\frac{\partial(\mathbf{r}^o/c^o)}{\partial \mathbf{s}^o} \quad \frac{\partial(\mathbf{r}^o/c^o)}{\partial \dot{\mathbf{s}}^o}\right],$$
$$\frac{\partial(\dot{\mathbf{r}}^o/c^o)}{\partial \boldsymbol{\beta}^o} = \left[\frac{\partial(\dot{\mathbf{r}}^o/c^o)}{\partial \mathbf{s}^o} \quad \frac{\partial(\dot{\mathbf{r}}^o/c^o)}{\partial \dot{\mathbf{s}}^o}\right].$$

Based on (1), (3) and (5), the partial derivatives in (52) can be calculated as follows:

1) $\frac{\partial(\mathbf{r}^o/c^o)}{\partial \mathbf{u}^o}$ is a $(M-1)\times 3$ matrix, and its $(i-1)$-th row is

$$\frac{\partial(r_{i1}^o/c^o)}{\partial \mathbf{u}^o} = \frac{1}{c^o}\left[\boldsymbol{\rho}^T_{\mathbf{u}^o,\mathbf{s}_i^o} - \boldsymbol{\rho}^T_{\mathbf{u}^o,\mathbf{s}_1^o}\right], \quad i = 2, 3, \ldots, M, \quad (53)$$

where $\boldsymbol{\rho}_{\mathbf{a},\mathbf{b}} \triangleq \mathbf{a}-\mathbf{b}/\|\mathbf{a}-\mathbf{b}\|$ denotes the unit vector from $\mathbf{b}$ to $\mathbf{a}$.

2) $\frac{\partial(\mathbf{r}^o/c^o)}{\partial \dot{\mathbf{u}}^o}$ is an $(M-1)\times 3$ zero matrix.

3) $\frac{\partial(\dot{\mathbf{r}}^o/c^o)}{\partial \mathbf{u}^o}$ is an $(M-1)\times 3$ matrix, and its $(i-1)$-th row is

$$\frac{\partial(\dot{r}_{i1}^o/c^o)}{\partial \mathbf{u}^o} = \frac{1}{c^o}\left(\left[\frac{(\dot{\mathbf{u}}^o-\dot{\mathbf{s}}_i^o)^T}{r_i^o} - \frac{(\mathbf{u}^o-\mathbf{s}_i^o)^T\dot{r}_i^o}{(r_i^o)^2}\right] - \left[\frac{[\dot{\mathbf{u}}^o-\dot{\mathbf{s}}_1^o]^T}{r_1^o} - \frac{(\mathbf{u}^o-\mathbf{s}_1^o)^T\dot{r}_1^o}{(r_1^o)^2}\right]\right), \quad i = 2, 3, \ldots, M. \quad (54)$$

4) $\frac{\partial(\dot{\mathbf{r}}^o/c^o)}{\partial \dot{\mathbf{u}}^o}$ is an $(M-1)\times 3$ matrix, and its $(i-1)$-th row is

$$\frac{\partial(\dot{r}_{i1}^o/c^o)}{\partial \dot{\mathbf{u}}^o} = \frac{1}{c^o}\left[\boldsymbol{\rho}^T_{\mathbf{u}^o,\mathbf{s}_i^o} - \boldsymbol{\rho}^T_{\mathbf{u}^o,\mathbf{s}_1^o}\right], \quad i = 2, 3, \ldots, M. \quad (55)$$

From (53), we observe that $\frac{\partial(\dot{\mathbf{r}}^o/c^o)}{\partial \dot{\mathbf{u}}^o} = \frac{\partial(\mathbf{r}^o/c^o)}{\partial \mathbf{u}^o}$.

5) $\frac{\partial(\mathbf{r}^o/c^o)}{\partial c^o}$ is an $(M-1)\times 1$ matrix, and its $(i-1)$-th row is

$$\frac{\partial(r_{i1}^o/c^o)}{\partial c^o} = -\frac{r_{i1}^o}{(c^o)^2}, \quad i = 2, 3, \ldots, M. \quad (56)$$

6) $\frac{\partial(\dot{\mathbf{r}}^o/c^o)}{\partial c^o}$ is an $(M-1)\times 1$ matrix, and its $(i-1)$-th row is

$$\frac{\partial(\dot{r}_{i1}^o/c^o)}{\partial c^o} = -\frac{\dot{r}_{i1}^o}{(c^o)^2}, \quad i = 2, 3, \ldots, M. \quad (57)$$

7) $\frac{\partial(\mathbf{r}^o/c^o)}{\partial \mathbf{s}^o}$ is an $(M-1)\times 3M$ matrix, and its $(i-1)$-th row is

$$\frac{\partial\left(r_{i1}^o/c^o\right)}{\partial \mathbf{s}^o} = \frac{1}{c^o}\left[\boldsymbol{\rho}_{\mathbf{u}^o,\mathbf{s}_1^o}^T, \mathbf{0}_{1\times 3(i-2)}, -\boldsymbol{\rho}_{\mathbf{u}^o,\mathbf{s}_i^o}^T, \mathbf{0}_{1\times 3(M-i)}\right],\quad i=2,3,\ldots,M. \tag{58}$$

8) $\frac{\partial(\mathbf{r}^o/c^o)}{\partial \dot{\mathbf{s}}^o}$ is an $(M-1)\times 3M$ zero matrix.

9) $\frac{\partial(\dot{\mathbf{r}}^o/c^o)}{\partial \mathbf{s}^o}$ is an $(M-1)\times 3M$ matrix, and its $(i-1)$-th row is

$$\frac{\partial\left(\dot{r}_{i1}^o/c^o\right)}{\partial \mathbf{s}^o} = \frac{1}{c^o}\left[\frac{\left(\dot{\mathbf{u}}^o-\dot{\mathbf{s}}_1^o\right)^T}{r_1^o} - \frac{\left(\mathbf{u}^o-\mathbf{s}_1^o\right)^T\dot{r}_1^o}{\left(r_1^o\right)^2}, \mathbf{0}_{1\times 3(i-2)}, -\left(\frac{\left(\dot{\mathbf{u}}^o-\dot{\mathbf{s}}_i^o\right)^T}{r_i^o} - \frac{\left(\mathbf{u}^o-\mathbf{s}_i^o\right)^T\dot{r}_i^o}{\left(r_i^o\right)^2}\right), \mathbf{0}_{1\times 3(M-i)}\right]. \tag{59}$$

10) $\frac{\partial(\dot{\mathbf{r}}^o/c^o)}{\partial \dot{\mathbf{s}}^o}$ is an $(M-1)\times 3M$ matrix and is equal to $\frac{\partial(\mathbf{r}^o/c^o)}{\partial \mathbf{s}^o}$.

## APPENDIX B
## THE DERIVATION OF (11)

According to the matrix inversion formula [36], we have

$$\begin{bmatrix}\mathbf{X}_{22} & \mathbf{X}_{23}\\ \mathbf{X}_{23}^T & \mathbf{X}_{33}\end{bmatrix}^{-1} = \begin{bmatrix}\mathbf{X}_{22}^{-1}+\mathbf{X}_{22}^{-1}\mathbf{X}_{23}\Gamma_1 X_{23}^T\mathbf{X}_{22}^{-1} & -\mathbf{X}_{22}^{-1}\mathbf{X}_{23}\Gamma_1\\ -\Gamma_1 X_{23}^T\mathbf{X}_{22}^{-1} & \Gamma_1\end{bmatrix},$$

where $\Gamma_1$ is defined in (13). Then, we can easily rewrite (10) into

$$CRLB_1(\boldsymbol{\theta}^o)^{-1} = \mathbf{X}_{11} - \mathbf{X}_{12}\mathbf{X}_{22}^{-1}\mathbf{X}_{12}^T + \left(\mathbf{X}_{13} - \mathbf{X}_{12}\mathbf{X}_{22}^{-1}\mathbf{X}_{23}\right)\Gamma_1\left(\mathbf{X}_{13} - \mathbf{X}_{12}\mathbf{X}_{22}^{-1}\mathbf{X}_{23}\right)^T. \tag{60}$$

Substituting (9) into $\mathbf{X}_{11} - \mathbf{X}_{12}\mathbf{X}_{22}^{-1}\mathbf{X}_{12}^T$, we obtain

$$\mathbf{X}_{11} - \mathbf{X}_{12}\mathbf{X}_{22}^{-1}\mathbf{X}_{12}^T = \left(\frac{\partial\boldsymbol{\alpha}^o}{\partial\boldsymbol{\theta}^o}\right)^T \mathbf{Q}_1^{-1}\left(\frac{\partial\boldsymbol{\alpha}^o}{\partial\boldsymbol{\theta}^o}\right), \tag{61}$$

where $\mathbf{Q}_1^{-1}$ is defined in (12). Similarly, we also have

$$\mathbf{X}_{13} - \mathbf{X}_{12}\mathbf{X}_{22}^{-1}\mathbf{X}_{23} = \left(\frac{\partial\boldsymbol{\alpha}^o}{\partial\boldsymbol{\theta}^o}\right)^T \mathbf{Q}_1^{-1}\left(\frac{\partial\boldsymbol{\alpha}^o}{\partial\boldsymbol{\beta}^o}\right). \tag{62}$$

Finally, with (61) and (62), (60) reduces to (11).

## APPENDIX C
## CALCULATION OF $\mathbf{G}_3$ AND $\mathbf{G}_4$

The calculation begins by defining $\mathbf{A}(i:j;m:n)$ as the submatrix formed by the $i$th to the $j$th rows and the $m$th to the $n$th columns of the matrix $\mathbf{A}$. using the definitions of $\mathbf{B}_1$, $\mathbf{G}_1$, $\mathbf{B}_2$, $\mathbf{G}_2$ and $\mathbf{B}_3$ and after some straightforward algebraic manipulations, we can express $\mathbf{G}_3$ as

$$\mathbf{G}_3(i-1;1:3) = \frac{1}{c^o}\left(\frac{\mathbf{u}^o-\mathbf{s}_i}{r_i^o} - \frac{\mathbf{u}^o-\mathbf{s}_1}{r_i^o}\right)^T - \frac{t_{i1}\boldsymbol{\varphi}_1(8)}{c^{o2}r_i^o\hat{r}_1^o}\left(\mathbf{u}^o-\mathbf{s}_1\right)^T,\quad i=2,\ldots,M, \tag{63}$$

$$\mathbf{G}_3(1:M-1;4:6) = \mathbf{0}_{(M-1)\times 3}, \tag{64}$$

$$\mathbf{G}_3(i-1;7) = -r_{i1}^2\Big/c^{o2}r_i^o,\quad i=2,\ldots,M, \tag{65}$$

$$\mathbf{G}_3(i+M-2;1:3) = \frac{1}{c^o}\left(\frac{\dot{r}_i^o(\mathbf{s}_i-\mathbf{s}_1)^T}{r_i^{o2}} - \frac{r_i^o(\dot{\mathbf{s}}_i-\dot{\mathbf{s}}_1)^T}{r_i^o}\right) + \left(\frac{\boldsymbol{\varphi}_1(8)}{c^{o2}\hat{r}_1^o}\left(\frac{\dot{r}_i^o t_{i1}}{r_i^{o2}} - \frac{\dot{t}_{i1}}{r_i^o}\right) + \frac{t_{i1}\hat{\dot{r}}_1^o\boldsymbol{\varphi}_1(8)}{c^{o2}r_i^o\hat{r}_1^{o2}}\right)\times\mathbf{1}_{1\times 3} - \frac{t_{i1}\boldsymbol{\varphi}_1(8)}{c^{o2}\hat{r}_1^o r_i^o}\left(\dot{\mathbf{u}}^o-\dot{\mathbf{s}}_1\right)^T,\quad i=2,\ldots,M, \tag{66}$$

$$\mathbf{G}_3(i+M-2;4:6) = \mathbf{G}_3(i-1;1:3),\quad i=2,\ldots,M, \tag{67}$$

$$\mathbf{G}_3(i+M-2;7) = \frac{\dot{r}_i^o t_{i1}^2}{r_i^{o2}} - \frac{2t_{i1}\dot{t}_{i1}}{r_i^o},\quad i=2,\ldots,M. \tag{68}$$

Similarly, $\mathbf{G}_4$ can be expressed as

$$\mathbf{G}_4(i-1;1:3M) = \frac{1}{c^o}\left[\frac{-\mathbf{d}_{i1}^T}{r_i^o}\quad \mathbf{0}_{1\times 3(i-2)}\quad \frac{(\mathbf{u}^o-\mathbf{s}_i)^T}{r_i^o}\quad \mathbf{0}_{1\times 3(M-i)}\right],\quad i=2,\ldots,M, \tag{69}$$

$$\mathbf{G}_4(1:M-1;3M+1:6M) = \mathbf{0}_{(M-1)\times 3M}, \tag{70}$$

$$\mathbf{G}_4(i+M-2;1:3M) = \frac{1}{c^o}\left[\frac{\mathbf{d}_{i1}^T\dot{r}_i^o}{r_i^{o2}} - \frac{\dot{\mathbf{d}}_{i1}^T}{r_i^o}, \mathbf{0}_{1\times 3(i-2)}, \frac{(\dot{\mathbf{u}}^o-\dot{\mathbf{s}}_i)^T}{r_i^o} - \frac{\dot{r}_i^o(\mathbf{u}^o-\mathbf{s}_i)^T}{r_i^{o2}}, \mathbf{0}_{1\times 3(M-i)}\right],\quad i=2,\ldots,M, \tag{71}$$

$$\mathbf{G}_4(M:2(M-1);3M+1:6M) = \mathbf{G}_4(1:M-1;1:3M). \tag{72}$$

Note that $\boldsymbol{\varphi}_1(8)$ is the estimate of $\eta_2(c^{o2})$ given by the first stage. Under small noise condition, we have the approximation $\varphi_1(8)\approx c^{o2}$, which indicates that the first stage solution is close to the true value. Directly making the approximations on the expressions (63)-(72) under the small Gaussian noise assumption and conditions (i-iii) gives

$$\mathbf{G}_3 \approx \left[\partial\boldsymbol{\alpha}^o/\partial\boldsymbol{\theta}^o\quad \partial\boldsymbol{\alpha}^o/\partial c^o\right],$$
$$\mathbf{G}_4 \approx -\partial\boldsymbol{\alpha}^o/\partial\boldsymbol{\beta}^o.$$

We conclude that the proposed solution can attain the CRLB accuracy approximately when the assumed conditions are satisfied.

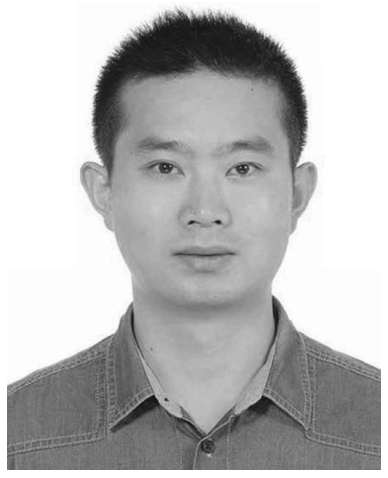

**BINGBING ZHANG** received the B.S. degree in electronic engineering and M.S. degree in electronic science and technology from the National University of Defense Technology (NUDT), Changsha, China, in 2011 and 2013, respectively, where he is currently pursuing the Ph.D. degree with the School of Electronic Science and Engineering. From 2016 to 2017, he visited the Department of Automation, Shanghai Jiao Tong University, China. His main research interests lie in the areas of underwater localization, navigation, and target tracking.

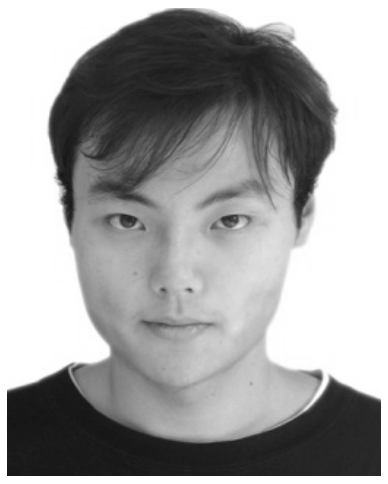

**YONGCHANG HU** was born in Xi'an, China, in 1988. He received the B.Sc. and M.Sc. degrees from Northwestern Polytechnical University, Xi'an, in 2010 and 2013, respectively, and the Ph.D. degree from the Circuits and Systems Group, Department of Microelectronics, Delft University of Technology, Delft, The Netherlands, in 2017. His research interests lie in the board area of signal processing and machine learning for communication and networking, particularly in localization/tracking, synchronization, OFDM, MIMO, mmWave communication, compressive sensing, modeling radio propagation channel, and random network analysis. He has been actively serving as a technical reviewer for several IEEE and EURASIP journals and major conferences.

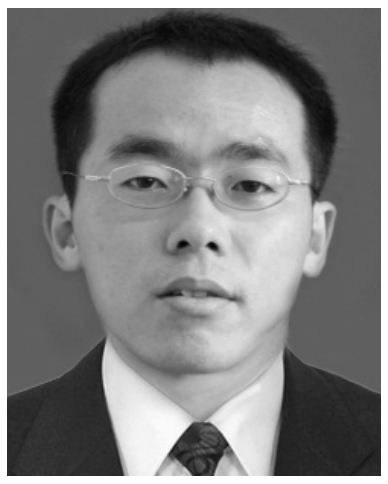

**HONGYI WANG** received the B.S. degree from the Beijing University of Aeronautics and Astronautics, Beijing, China, in 2001, and the M.S. and Ph.D. degrees in electronic engineering from the National University of Defense Technology, Changsha, China, in 2003 and 2008, respectively. He is currently an Assistant Professor with the ASIC Research and Development Center, National University of Defense Technology. His research interests include digital design, wireless system design, test, and debug.

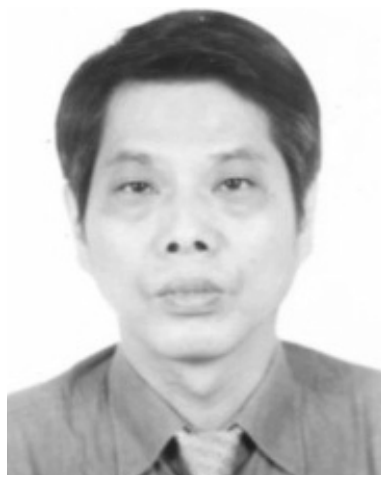

**ZHAOWEN ZHUANG** received the B.S. and M.S. degrees from the National University of Defense Technology (NUDT), Changsha, China, in 1981 and 1984, respectively, and the Ph.D. degree from the Beijing Institute of Technology, Beijing, China, in 1989. He is currently a Professor and the Deputy Head of NUDT. His research interests lie in the areas of artificial intelligence and target recognition.